\documentclass[usenatbib]{mn2e}
\usepackage{epsfig}
\usepackage{subfigure}
\usepackage{amsmath,amssymb}
\usepackage{natbib}
\usepackage{color}
\usepackage{rotating}
\usepackage{braket}
\begin{document}

\title[Molecular hydrogen emission in the LMC]
{Molecular hydrogen emission in the interstellar medium of the Large Magellanic Cloud}
\author[Naslim, N., et al.]{Naslim~N.$^{1}$\thanks{E-mail: naslimn@asiaa.sinica.edu.tw}, F.~Kemper$^{1}$\thanks{E-mail: ciska@asiaa.sinica.edu.tw}, S.~C.~Madden$^2$, S.~Hony$^3$, Y.-H.~Chu$^1$, F.~Galliano$^2$, C.~Bot$^4$,  
 \newauthor
Y.~Yang$^{1}$,
J.~Seok$^{1}$,
J.~M.~Oliveira$^5$,
J.~Th.~van~Loon$^5$,
 M.~Meixner$^6$, A.~Li$^{7}$,    
\newauthor
A.~Hughes$^3$,
K.~D.~Gordon$^6$,
M.~Otsuka$^1$,
H. Hirashita$^{1}$,
O.~Morata$^{1}$,
V.~Lebouteiller$^2$,
\newauthor
R.~Indebetouw$^8$,
S.~Srinivasan$^1$,
 J.-P.~Bernard$^9$, W.~T.~Reach$^{10}$ \\
        $^1$Academia Sinica Institute of Astronomy and Astrophysics,  Taipei 10617, Taiwan R.O.C\\
        $^2$Laboratoire AIM, CEA/DSM - CEA Saclay, 91191 Gif-sur-Yvette, France \\
        $^3$Max-Planck-Institute for Astronomy, K\"onigstuhl 17, 69117 Heidelberg, Germany \\
        $^4$Observatoire astronomique de Strasbourg, Universit\'e de Strasbourg, \\
        CNRS, UMR 7550, 11 rue de l’Universit\'e, F-67000 Strasbourg, France \\
        $^{5}$Astrophysics Group, Lennard Jones Laboratories, Keele University, ST5 5BG, UK\\     
        $^6$Space Telescope Science Institute, 3700 San Martin Drive, Baltimore, MD 21218, USA\\
        $^{7}$Department of Physics and Astronomy, University of Missouri, Columbia, MO 65211, USA\\
        $^8$Department of Astronomy, University of Virginia, PO Box 400325, VA 22904, USA\\
        $^9$CNRS, IRAP, 9 Av. colonel Roche, BP 44346, F-31028 Toulouse Cedex 4, France \\
        $^{10}$USRA-SOFIA Science Center, NASA Ames Research Center, Moffett Field, CA 94035, USA   \\}

\date{Accepted .....
      Received ..... ;
      in original form .....}

\pagerange{\pageref{firstpage}--\pageref{lastpage}}
\pubyear{2014}
\label{firstpage}
\maketitle

\begin{abstract}

We present the detection and analysis of molecular hydrogen emission
toward ten interstellar regions in the Large Magellanic Cloud.  We
examined low-resolution infrared spectral maps of twelve regions
obtained with the {\it Spitzer} infrared spectrograph (IRS). The pure
rotational 0--0 transitions of H$_2$ at 28.2 and 17.1${\,\rm \mu m}$
are detected in the IRS spectra for ten regions. The higher level
transitions are mostly upper limit measurements except for three
regions, where a 3$\sigma$ detection threshold is achieved for lines
at 12.2 and 8.6${\,\rm \mu m}$. The excitation diagrams of the
detected H$_2$ transitions are used to determine the warm H$_2$ gas
column density and temperature. The single-temperature fits through the
lower transition lines give temperatures in the range $86-137\,{\rm
K}$. The bulk of the excited H$_2$ gas is found at these temperatures
and contributes $\sim5-17$\,per\,cent to the total
gas mass. We find a tight correlation of the H$_2$ surface brightness with polycyclic aromatic hydrocarbon
and total infrared emission, which is a clear
indication of photo-electric heating in photodissociation regions. We
find the excitation of H$_2$ by this process is equally efficient in
both atomic and molecular dominated regions. We also present the
correlation of the warm H$_2$ physical conditions with dust
properties. The warm H$_2$ mass fraction and excitation temperature
show positive correlations with the average starlight
intensity, again supporting H$_2$ excitation in
photodissociation regions.

\end{abstract}


\section{Introduction}

Molecular hydrogen is a major component of the interstellar medium
(ISM) in gas-rich star-forming galaxies. Quantifying the reservoir of
molecular gas and its distribution and physical conditions is
essential to understanding the process of conversion of gas into
stars. The bulk of molecular gas in the form of cold H$_2$ remains
undetectable due to the lack of permanent dipole moment of
H$_2$. Therefore, to quantify the molecular gas reservoir, we usually
rely on the second most abundant molecule, CO, using a CO-to-H$_2$
conversion factor X$_{\rm co}$ \citep{bolatto13}.  However, CO can
have difficulties in tracing all of the molecular gas under certain
conditions, e.g. low extinction, when substantial H$_2$ may exist
outside of the CO-emitting region \citep{wolfire10}.  

 The pure rotational 0--0 transitions of H$_2$ due to the molecule's
quadrupole moment are more direct tracers of the H$_2$ gas, although
they are only excited at higher temperatures than those prevalent
in molecular cloud interiors. These mid-infrared transitions trace the
bulk of the warm molecular gas with temperatures between 100 and
1000\,K, which is a small but non-negligible fraction of the total
molecular gas reservoir.  

The major excitation mechanisms of H$_2$ in star-forming galaxies are
thought to be the far-ultraviolet (FUV) radiation from massive stars
in photodissociation regions (PDRs) \citep{hollenbach97, tielens93} or
collisional excitation in shocks \citep{draine83}. PDRs are formed at
the edges of molecular clouds where an incident FUV radiation field
controls the physical and chemical properties of the gas. The UV
photons can dissociate molecules like H$_2$ and CO, and further ionize
the atoms, causing a stratified chemical structure with partially
ionized hydrogen, neutral hydrogen, molecular hydrogen and CO gas.
Moreover, photo-electric heating will occur where electrons ejected
from polycyclic aromatic hydrocarbons (PAHs) and dust grains due to incident UV photons collide with the
ambient gas, and transfer the excess kinetic energy to the molecules
and atoms. Consequently, emission due to warm molecular H$_2$ is
expected to arise at the edge of the molecular clouds, at the
innermost part of the PDR, adjacent to the region from which the PAH
emission arises \citep{tielens93}.

{\it Spitzer} observations revealed
rotational H$_2$ emission in extragalactic objects with various
physical mechanisms responsible for the H$_2$ excitation. For example,
using the {\it Spitzer} Infrared Nearby Galaxy Survey (SINGS;
\citealt{kennicutt03}), \citet{roussel07} showed that
PDRs at the interface between ionized
H{\,\sc ii} regions and dense molecular clouds are the major source
for the H$_2$ emission in normal star-forming galaxies. In addition,
the excess H$_2$ emission relative to hydrogen recombination lines and
PAH emission observed in spiral
galaxies \citep{beirao09}, ultraluminous infrared galaxies
\citep{higdon06} and active galactic nuclei \citep{ogle10} are known
to be shock excited. \citet{ingalls11} suggested that mechanical
  heating via shocks or turbulent dissipation is the dominant H$_2$
  excitation source in non-active galaxies. This argument has also been invoked in \citet{beirao12} to explain H$_2$ emission in the
  starburst ring of Seyfert 1 galaxy NGC 1097.

The uncertainty of the amount of H$_2$, physical properties and
excitation conditions in galaxies are exacerbated due to the
relatively weak detection threshold of H$_2$
\citep{habart05}. \citet{hunt10} presented global mid-infrared spectra
of several nearby low-metallicity dwarf galaxies. They detected an
excess (6\,per\,cent) amount of H$_2$ to PAH emission in many dwarf
galaxies compared to SINGS normal galaxies. This excess could arise
due to PAH deficit in metal-poor galaxies. In metal-poor environments,
due to the diminished dust shielding, FUV radiation from hot stars
penetrates deeper into the surrounding dense ISM and significantly
affects both physical and chemical processes by the photodissociation
of most molecules including CO, except molecular hydrogen H$_2$ (H$_2$
survives due to self-shielding) \citep{madden97, wolfire10}.

The nearby Large Magellanic Cloud (LMC) galaxy is an excellent site to
explore the ISM H$_2$ emission due to its proximity ($\sim$
$49.97\,{\rm kpc}$; \citealt{pietrzy13}) and low extinction along the
line of sight. In this paper we present, for the first time, the
detection of H$_2$ rotational transitions in the LMC in {\it Spitzer}
spectra of some selected ISM regions, that were obtained as part of
Surveying the Agents of Galaxy Evolution (SAGE-LMC) spectroscopic
program (SAGE-spec) \citep{meixner10, kemper10}. We quantify the
excitation temperature and mass of the warm H$_2$ using excitation
diagram analysis and compare the mass estimates with total gas content
deduced from CO, H{\,\sc i} and H$\alpha$ observations. We examine
correlations between the warm H$_2$ emission, PAH, mid- and total
infrared continuum emission to investigate the nature and dominant
excitation mechanism of the warm H$_2$ gas in the LMC. We compare our
LMC results to other recent analyses of the warm
H$_2$ emission in normal star-forming and dwarf galaxy samples.

\section{Observations and data reduction}

\begin{figure*}
\centering 
\epsfig{file=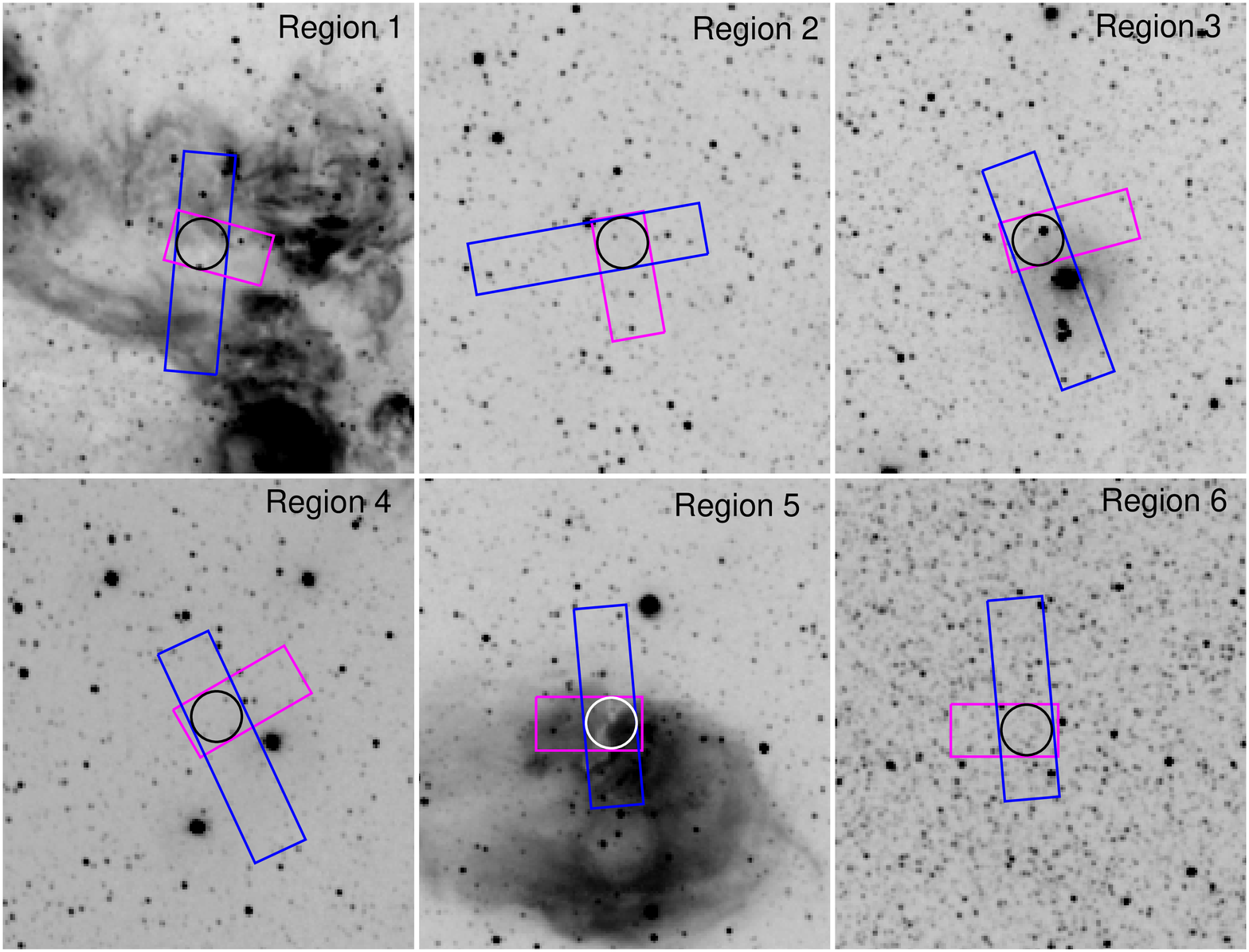, scale=0.4}
\epsfig{file=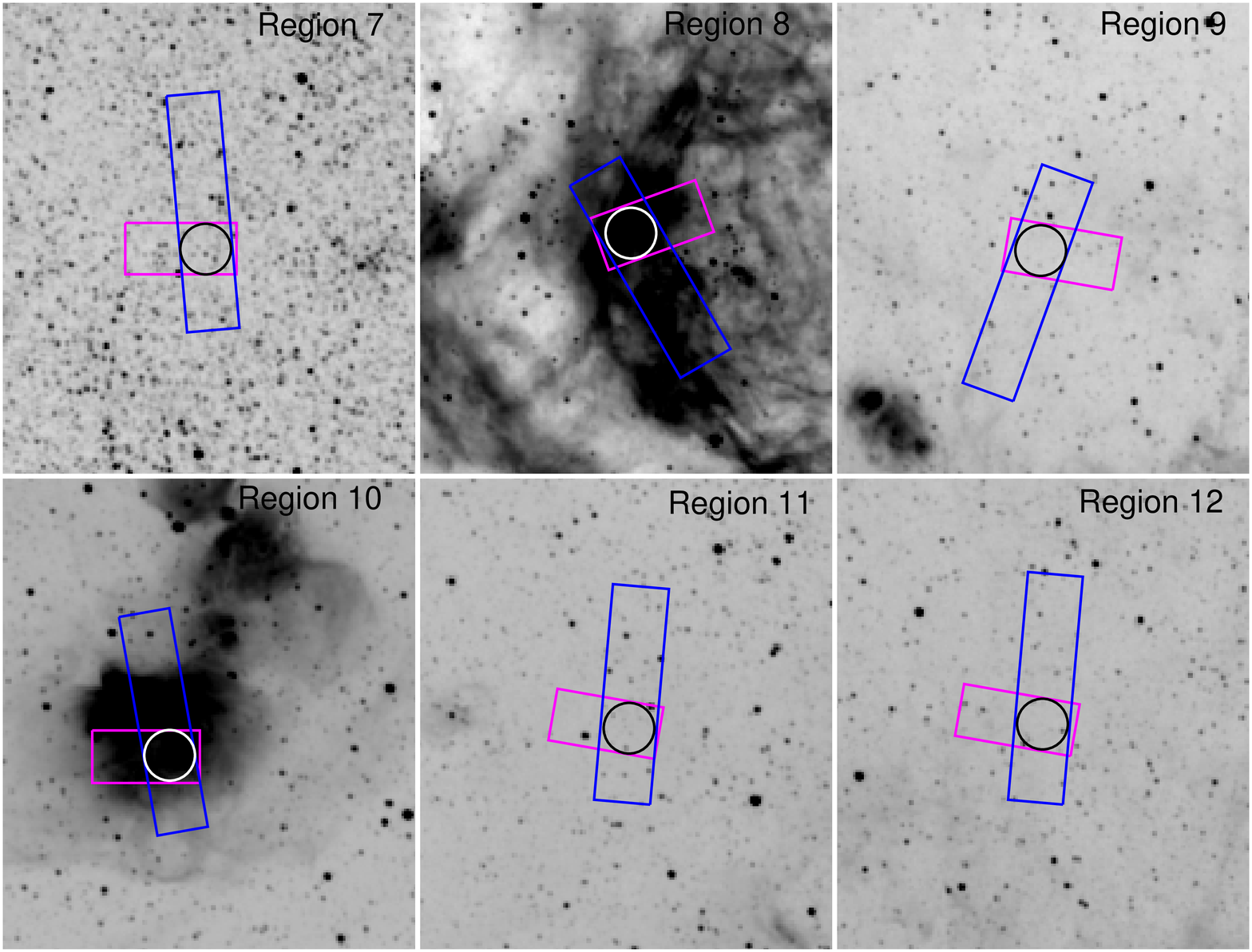, scale=0.4}
\caption{Observed ISM regions are shown on the MCELS H$\alpha$
  images. Overlayed rectangles are the coverage for IRS Short Low
  (small rectangle) and IRS Long Low (long rectangle) slits. The overlap region is
  1$\times$1\,arcmin$^2$ and the circles (30\,arcsec radii) show
  the regions over which the spectra are integrated.}
\label{LMCregions}
\end{figure*}

For our analysis we used photometric and spectroscopic data obtained
with the {\it Spitzer} and {\it Herschel} legacy programs. As part of the SAGE-LMC,
photometric images have been taken using IRAC (InfraRed Array Camera; 3.6, 4.5, 5.8,
8.0${\,\rm \mu m}$) and MIPS (Multiband Imaging Photometer for
 {\it Spitzer}; 24, 60, 100 and 160${\,\rm \mu m}$) on
board the {\it Spitzer} Space Telescope. As a
follow up to SAGE-LMC, the spectroscopic survey (SAGE-spec), used the IRS
on {\it Spitzer} to obtain spectral maps of a total of 20 extended regions,
which were grouped into 10 H{\,\sc ii} regions and 10 diffuse regions
\citep{kemper10}. These regions were selected based on their infrared
colors and other characteristics with a goal to sample a wide range of
physical conditions. Spectra were taken in mapping mode with the
low-resolution modules Short Low 2 (SL2), Short Low 1 (SL1), Long Low
2 (LL2) and Long Low 1 (LL1) over a wavelength range $5-38{\rm \mu
 m}$. Our analysis included the 10 regions which were grouped
into diffuse regions in the SAGE-spec samples (regions 1--10 in Table \ref{coord}) and two additional atomic
regions in the LMC from PID\,40031, that were observed in similar way (PI: G. Fazio; regions 11 and
12). Further details of sample selection based on infrared colors can
be found in \citet{kemper10}. When we examined these 12 regions on an
H$\alpha$ image from the Magellanic Cloud Emission Line Survey
(MCELS; \citealt{smith98}), two regions are found to be right inside H{\,\sc
 ii} regions and another two are placed at the edge of H{\,\sc ii}
regions. We will discuss the spectra of these regions in section 3. 
Fig. \ref{LMCregions} shows the outlines of the twelve spectral maps on top of cut-outs of the MCELS
H$\alpha$ image. The coordinates and associated H{\,\sc ii} regions are given in Table \ref{coord}.

\begin{table}
\caption{Coordinates of our sample ISM regions in the LMC.}
\begin{center}
\begin{tabular}{cccc}
\hline
Region & RA (J2000)  & Dec (J2000) & H\,{\sc ii} regions\\
\hline
   1    &05h32m02.18s    & -68d28m13.6s & N\,148\\
   2    &05h43m42.01s  &   -68d15m07.4s&  \\
   3    &05h15m43.64s  &   -68d03m20.3s & \\
   4    &04h47m40.85s  &   -67d12m31.0s & \\
   5    &05h55m54.19s  &   -68d11m57.1s & N\,75\\
   6    &05h47m16.29s  &    -70d42m55.5s & \\
   7    &05h35m09.36s  &    -70d03m24.2s & \\
   8    &05h26m25.17s  &    -67d29m08.1s & N\,51 \\
  9    &05h32m10.73s  &    -68d21m10.8s & \\
   10    & 05h32m22.95s &    -66d28m41.5s & N\,55\\
   11    &05h31m07.10s  &    -68d19m12.0s &\\
   12    &05h43m39.65s  &    -68d46m18.6s &\\

\hline
\end{tabular}
\end{center}
\label{coord}
\end{table}

\begin{figure*}
\centering 
\includegraphics[angle = 0]{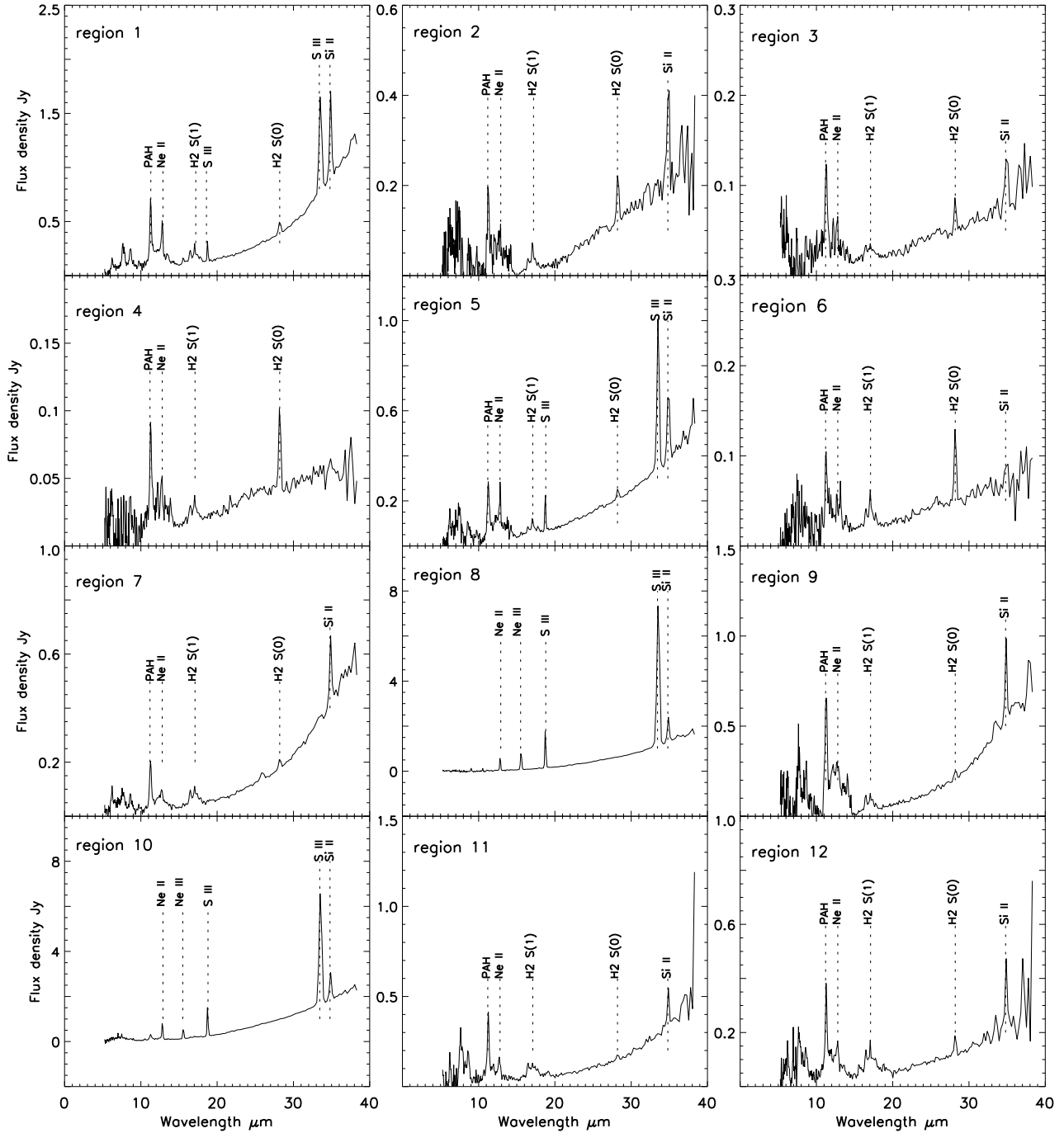}
\caption{The integrated IRS spectra for all 12 sample regions. The
  pure rotational transitions of molecular hydrogen S(0) and S(1) as
  well as several atomic lines and 11.3${\,\rm \mu m}$ PAH feature are
  indicated with dashed lines.}
\label{Spec}
\end{figure*}

The data reduction pipeline at the {\it Spitzer} Science Center was
used to reduce and process the raw data products. The details of data
reduction method, background subtraction, bad pixel masking and
spectral extraction can be found in \citet{kemper10}. Using the IDL
package CUBISM \citep{smith07} the individual spectral orders were
combined into spectral cubes and custom software
\citep{sandstrom09} was used to merge those spectral cubes. The
individual extracted single order spectra (6 for IRS) were merged into
a single spectrum using a combination of the overlap regions. The
overlap regions in wavelength were used to adjust the flux levels of the spectral
orders, except in cases where the S/N of the spectra did not allow
for an accurate measurement of the offset in the overlap regions. The
background subtraction was made using the dedicated observations of
the off-LMC background region. As can be seen in Fig. \ref{LMCregions}, the area of overlap between SL and LL slits
covers a region of 1$\times$1\,arcmin$^2$. The orientation
of SL and LL slit apertures is marked on MCELS H$\alpha$ image in
Fig. \ref{LMCregions}. The merged spectra were spatially integrated
over a circle of radius 30$\,$arcsec within the overlapping area of
1$\times$1\,arcmin$^2$. The
overall level of the IRS spectrum in SL and LL was scaled to
the photometric fluxes obtained from IRAC 8.0${\,\rm \mu m}$ and MIPS
 24${\,\rm \mu m}$, respectively. The photometry was
extracted from the latest set of full SAGE-LMC mosaics using the same
apertures as were used for the spectroscopy (subtracting the off-LMC
background region where appropriate). In general, the MIPS 24${\,\rm
  \mu m}$ was used to set the overall level of the merged IRS
spectrum, supplemented with the IRAC 8.0${\,\rm \mu m}$ data point
when it did not impose too strong an offset between the SL and LL
orders. A S/N of $\ge$10 was achieved for the integrated spectra of
each region. The integrated spectra of all twelve regions are shown
in Fig. \ref{Spec}.

To study the associated dust properties of these regions,
 we supplement the analysis with the SAGE-LMC and the {\it
  Herschel} Inventory of The Agents of Galaxy Evolution (HERITAGE)
photometric data \citep{meixner13}. For the HERITAGE project, the
LMC has been surveyed using the Photodetector Array Camera and
Spectrometer (PACS; 100 and 160${\,\rm \mu m}$) and the Spectral and
Photometric Imaging Receiver (SPIRE; 250, 350, and 500${\,\rm \mu
  m}$). The details of the observations, data reduction and data processing
can be found in \citet{meixner10}. To extract fluxes we carried out
aperture photometry using the aperture photometry tool (APT) and
compared the results with the APER routine in IDL. The APT tool performs photometry
by summing all pixels within a circular or elliptical aperture. We
used a circular aperture of radius 30$\,$arcsec for all twelve sample
regions. A background flux of the same aperture size was determined from
an off-LMC region, since we want to avoid any contamination of the
emission to the background noise. The error on the flux densities comes from
source aperture measurements which include background noise obtained
by standard deviation of pixels within the sky aperture, and flux
calibration errors. To determine the overall uncertainty on the flux
density, we quadratically added the flux extraction error and flux
calibration errors which are 5\,per\,cent for MIPS, 15\,per\,cent for PACS and 7\,per\,cent
for SPIRE data \citep{meixner13}.

\begin{figure*}
\centering 
\includegraphics[trim=0.0cm 0.50cm 0.0cm 0.0cm, clip = true,angle = 0]{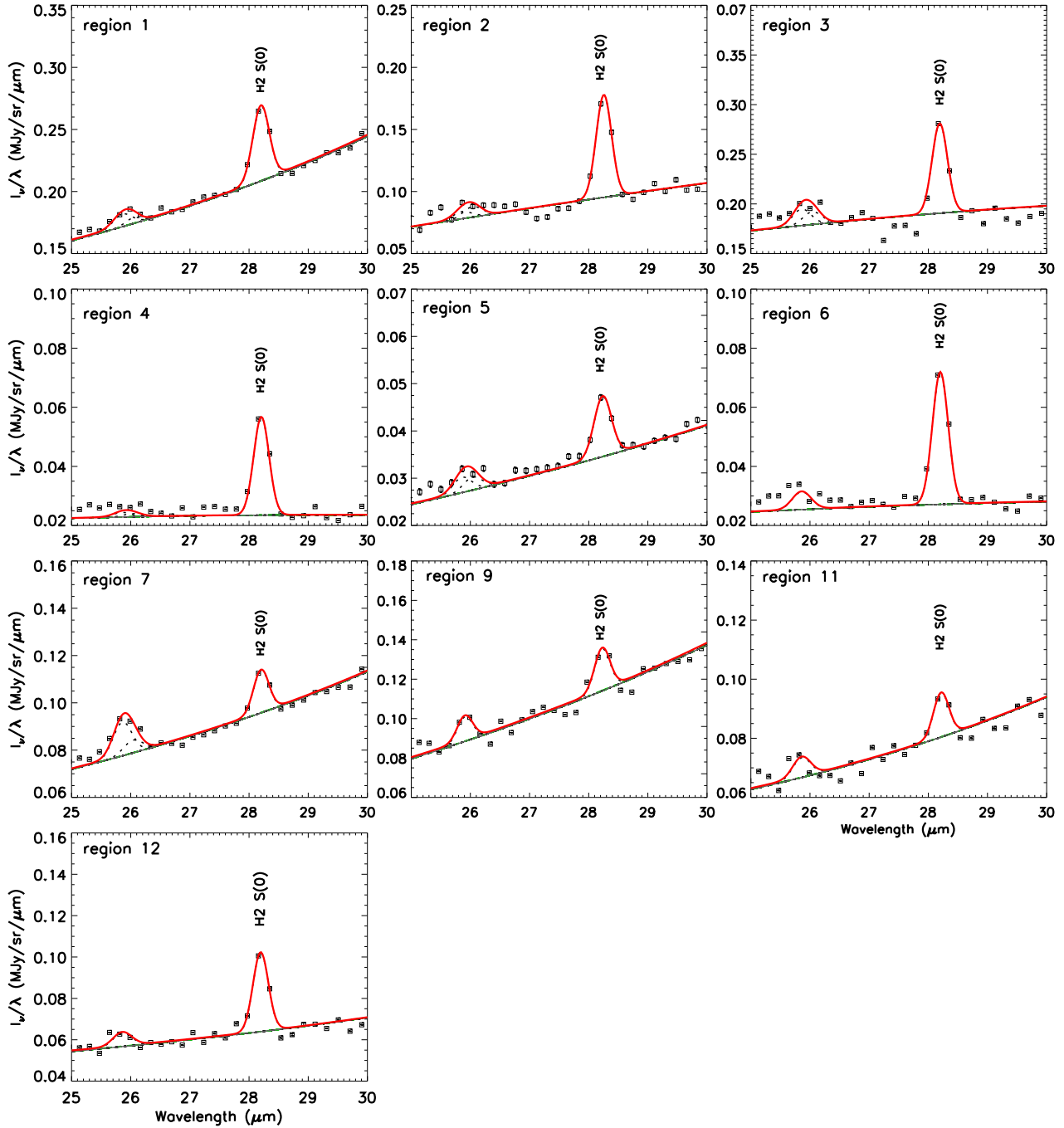}
\caption{Segments of H$_2$ S(0) line fits using PAHFIT. The
  black squares indicate the observed spectra and the best fit model is shown in solid line.}
\label{pahfit}
\end{figure*}
\begin{figure*}
\centering 
\includegraphics[trim=0.0cm 0.50cm 0.0cm 0.0cm, clip = true,angle = 0]{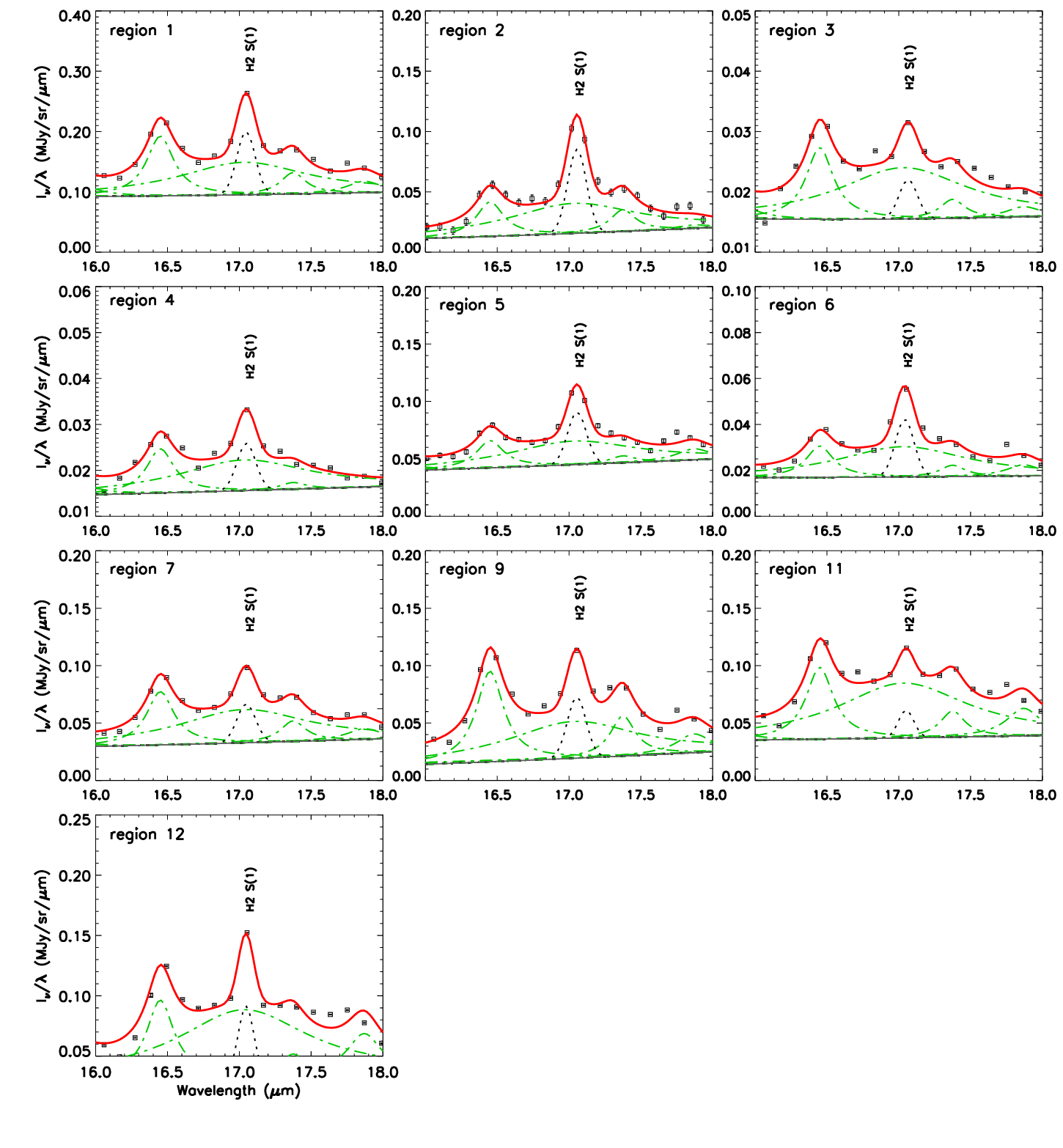}
\caption{Segments of H$_2$ S(1) line fits using PAHFIT. The
  black squares indicate the observed spectra, the superimposed are:
  the best fit model (solid), fits to PAH features (dash dot) and the H$_2$ lines
   (dotted).}
\label{pahfit2}
\end{figure*}
\begin{figure*}
\centering 
\includegraphics[trim=0.0cm 14.0cm 0.0cm 0.0cm, clip = true, angle = 0]{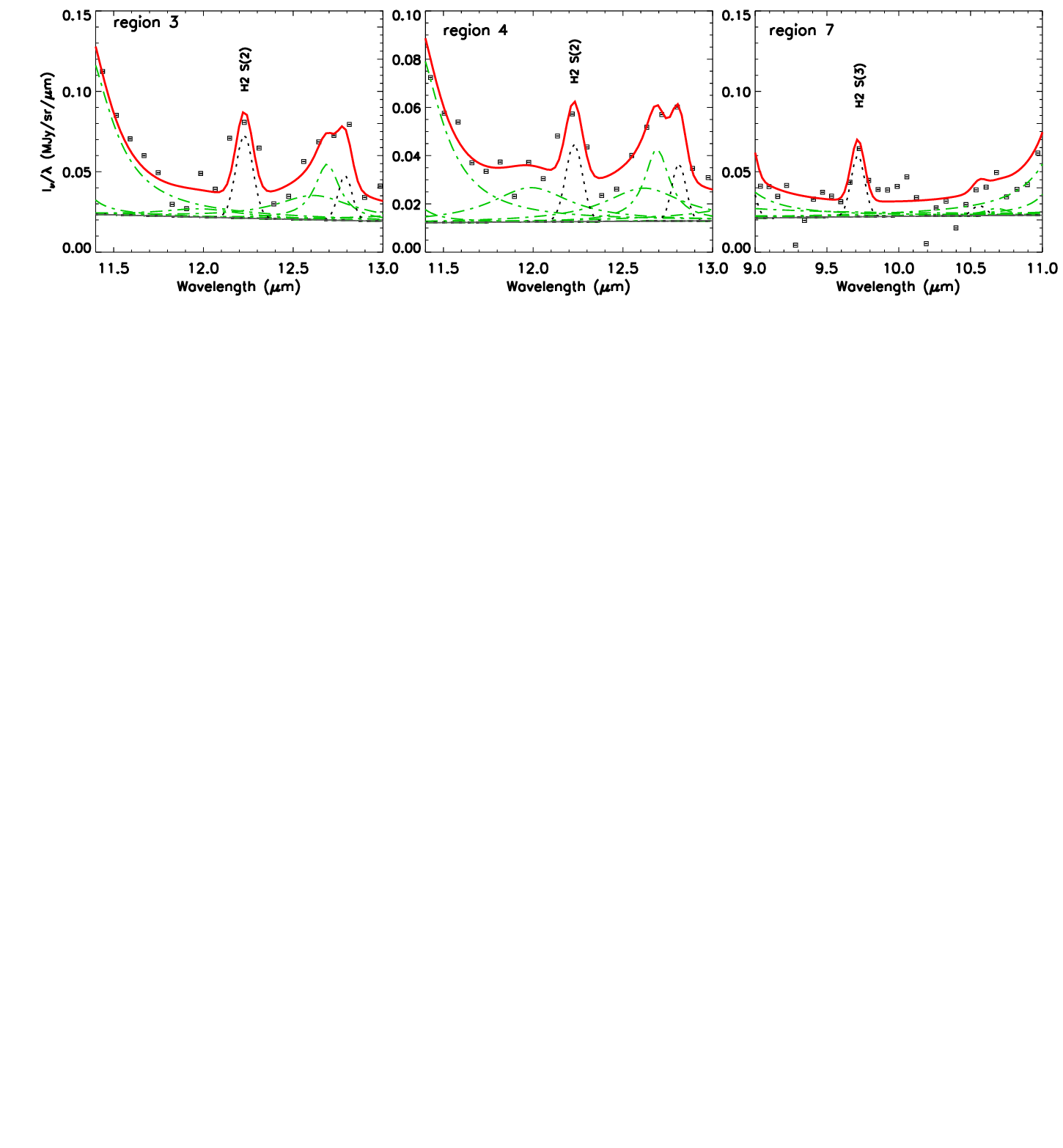}
\caption{Segments of H$_2$ S(2) and S(3) line fits using PAHFIT. The
  black squares indicate the observed spectra, the superimposed are:
  the best fit model (solid), fits to PAH features (dash dot) and the H$_2$ lines
   (dotted).}
\label{pahfit3}
\end{figure*}

\section{Spectral features}

The integrated spectra were decomposed using IDL package PAHFIT
\citep{smith07} which can be used to simultaneously fit the PAH
features, dust continuum, molecular lines, and fine structure
lines. The default PAHFIT package returns only the formal statistical
uncertainties for fitted parameters using the given flux
uncertainties obtained from the IRS pipeline. We applied a random
perturbation method to the existing PAHFIT package
to treat better uncertainties for line strengths. We randomly
perturbed the flux at each wavelength in a spectrum for a
user-specified number of times (100) taking a Gaussian distribution
for given standard deviation ($\sigma$). Here, $\sigma$ is taken as
quadratic sum of flux uncertainties obtained from IRS pipeline and
standard deviation ($\sigma_i$) to the continuum. $\sigma_i$ is
determined by fitting a polynomial to a user specified region of
continuum where no lines are present. We finally applied PAHFIT to
each of these perturbed spectra to obtain the distribution of best-fit
parameters.

The low-resolution IRS spectra allow us to detect the rotational
transition lines of H$_2$, the aromatic band features due to PAH and
the fine structure lines of [Ne\,{\sc ii}] (12.8${\,\rm \mu m}$),
[S{\,\sc iii}] (18.7, 33.4${\,\rm \mu m}$) and [Si{\,\sc ii}]
(34.9${\,\rm \mu m}$). Ten out of the twelve regions showed strong H$_2$
S(0) ($v$=0-0, J=2-0, 28.2${\,\rm \mu m}$) and S(1) ($v$=0-0, J=3-1,
17.0${\,\rm \mu m}$) lines. These lines are detected with a
$\sim$5$\sigma$ amplitude sensitivity. At shorter wavelengths
(5-13${\,\rm \mu m}$) the upper level transitions S(2), S(3), S(4),
S(5), S(6) and S(7) (S(2): $v$=0-0, J=4-2, 12.2${\,\rm \mu m}$, S(3):
$v$=0-0, J=5-3, 9.6${\,\rm \mu m}$, S(4): $v$=0-0, J=6-4, 8.0${\,\rm
  \mu m}$, S(5): $v$=0-0, J=7-5, 6.9${\,\rm \mu m}$, S(6): $v$=0-0,
J=8-6, 6.1${\,\rm \mu m}$, S(7): $v$=0-0, J=9-7, 5.5${\,\rm \mu m}$)
are highly contaminated by aromatic band features of PAHs at 6.2, 7.7,
8.6, 11.3 and 12.6${\,\rm \mu m}$. Fig. \ref{pahfit}, \ref{pahfit2} and \ref{pahfit3} show the best
line fits for all detected H$_2$ lines. In regions 3 and 4 the S(2)
transition and in region 7 the S(3) transition are strong enough to
measure the equivalent widths. These lines are detected with a line
strengths of 3$\sigma$. These line fits are also shown in
Fig. \ref{pahfit}. A 2$\sigma$ upper limit to the integrated intensity
was estimated for all remaining non-detected H$_2$ lines. To determine
the upper limit intensity, we first subtracted the fit to all PAH
features and dust continuum from the observed spectrum. Then we
calculated, the root mean square of the residual spectrum at the line
within a wavelength range $\lambda_0\pm4\times$FWHM$_{\lambda}$, and
the integrated intensity \citep{ingalls11}. The measured line intensities of detected H$_2$ lines 
and upper limits of non-detections are given in Table \ref{lines}.

The integrated IRS spectra of regions 1, 5, 8 and 10 show strong ionic lines due
to [S{\,\sc iii}] 18.7, and 33.4${\,\rm \mu m}$, whereas these lines
are not detectable in the remaining eight regions. The [Ne{\,\sc iii}]
15.1${\,\rm \mu m}$ line is detected in regions 8 and 10 while there
is no detection of H$_2$ rotational transitions in these
regions. These strong ionic lines indicate the presence of associated
H{\,\sc ii} regions, hence we examined the environments of those
regions using the H$\alpha$ map obtained from MCELS which has an angular
resolution 3$\,$arcsec. Region 1 is located at the inner edge of
northern lobe of H{\,\sc ii} region N\,148\,C in the supergiant shell
LMC 3 \citep{book08} which is at the northwest of 30\,Doradus. Region
5 is located at the northern edge of H{\,\sc ii} region N\,75 and
region 8 is inside N\,51 \citep{bica99}. Region 10 is located inside
the southernmost lobe of H{\,\sc ii} region N\,55 in the supergiant shell LMC
4 \citep{dopita85, book08} which is at the northern edge of the
LMC. All remaining eight regions appear to be isolated ISM regions in
the H$\alpha$ image, where [Ne\,{\sc iii}] and [S\,{\sc iii}] lines
are not detected, while the [Si\,{\sc ii}] 34.2${\,\rm \mu m}$ and
[Ne\,{\sc ii}] 12.8${\,\rm \mu m}$ are detected in all the 12
regions. 

Our analysis of the regions using H$_2$ rotational lines, H{\,\sc
  i}, H$\alpha$, and CO data will allow us to find out whether these
are either molecular or atomic dominated regions or ionized regions. A
detailed analysis of spatial distribution of ionic and molecular
species of some of these regions along with PDR analysis will be
presented in a future paper. In this paper we focus on the properties
of H$_2$ emissions detected toward ten regions (Regions 1--7, 9, 11
and 12) discussed above. We have not included regions 8 and 10 in rest
of our analysis. In sections 6 and 7 we will also discuss the
molecular, atomic and ionized gas contents and physical
parameters of dust in these regions. We discuss the analysis of these regions
with notations as numbers shown in Fig. \ref{Spec} and Table
\ref{coord}.

\begin{figure*}
\centering 
\includegraphics[angle = 0]{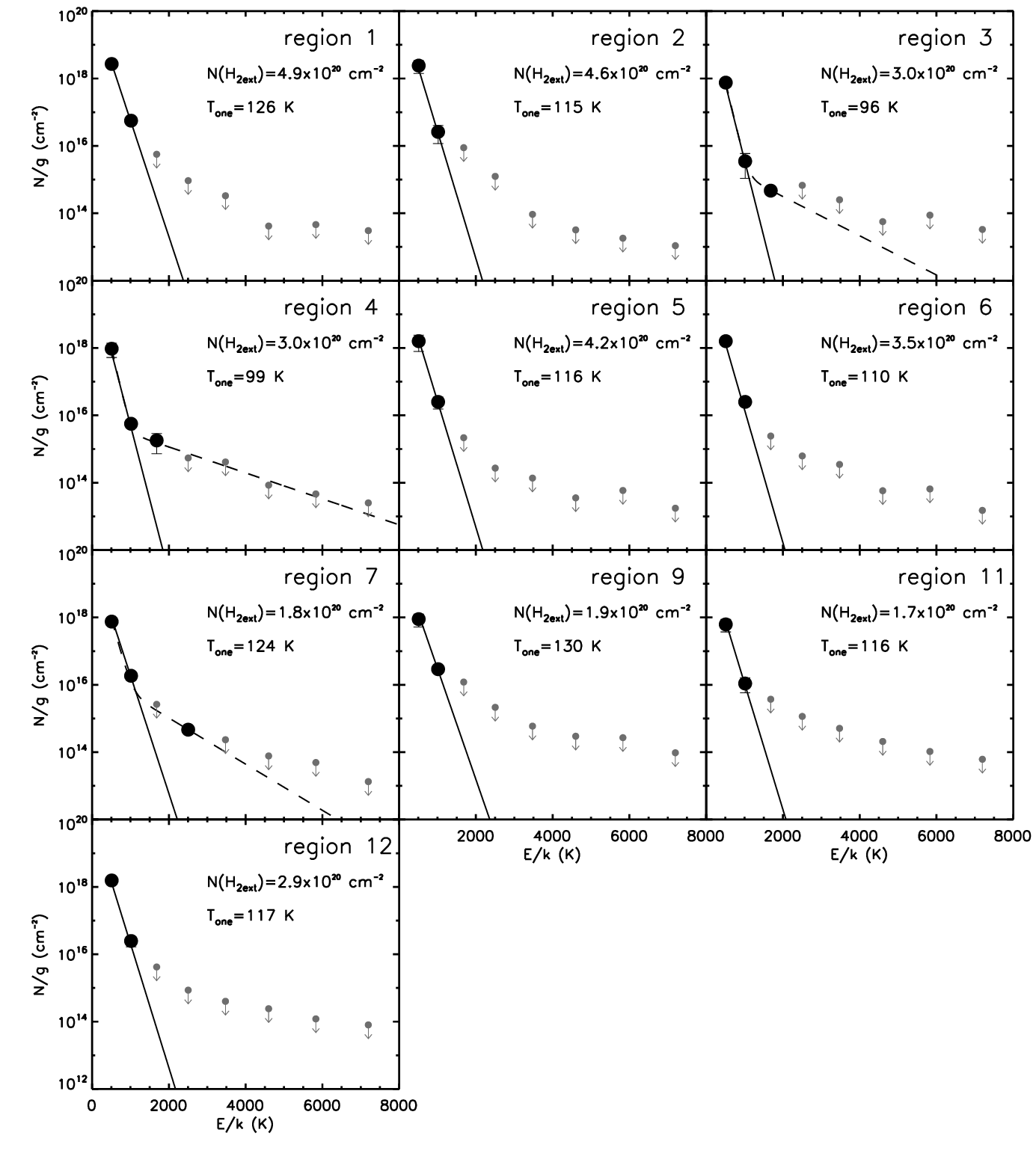}
\caption{Excitation diagrams of H$_2$ for ten regions where the lines
  are clearly detected. The black dots show detected transitions
  and the gray dots with upper limits are non-detections. The solid
  black line shows the single-temperature model fit and dashed lines for
  regions 3, 4 and 7 are two-temperature fits. The total
  excited H$_2$ column densities N(H$_{2\,\rm ext}$) and excitation temperature T$_{\rm one}$ from the
 single-temperature fits are shown in each panel.}
\label{popdiag}
\end{figure*}

\section{H$_2$ column density and excitation temperature}

The excitation diagram of H$_2$ rotational emission can be
used to derive the level populations and thus excitation temperature of warm
molecular hydrogen gas. Using the measured line intensities
$I_{\rm obs(i, j)}$ of each transition we derived the H$_2$ column
densities from the following formula:
\begin{equation}
N_{\rm obs(i,j)} = \frac{4\pi \lambda I_{\rm obs(i, j)}}{hcA_{\rm (i,j)}},
\end{equation}
where $A_{(i,j)}$ are the
Einstein coefficients. We assume that the lines are optically thin, the radiation is
isotropic and rotational levels of H$_2$ are thermalized. Under these
assumptions the level population follow the Boltzmann distribution law
at a given temperature and the total H$_2$ column densities
$N(\rm H_{2\,\rm ext})$ can be determined in the Local Thermodynamic Equilibrium
(LTE) condition using the formula:  

\begin{equation}
N_{\rm j}/g_{\rm j} = \frac{N(\rm H_{2\,{\rm ext}})exp^{-\it E_{\rm j}/kT}}{Z(T)},
\end{equation}

where $Z(T)\,\sim\,0.0247T/[1-\rm exp(-6000\,\rm K/{\it T})]$ is the partition function \citep{herbst96}. 
 
This function defines a straight line in the excitation diagram with a
slope $1/T$, ie. the upper level column densities normalized with the
statistical weight in logarithmic scale as a function of the upper
level energy \citep{smith99}. The statistical weight is $g_{\rm
    j}=(2s+1)(2j+1)$, with spin number $s=0$ for even j and
  $s=1$ for odd j \citep{rosenthal00, roussel07}. In the LTE condition
  with gas temperature greater than 300\,K, an equilibrium ortho-to-para ratio
  of 3 is usually assumed. However, for our analysis we adopt an ortho-to-para
  ratio of 1.7 which is appropriate for gas temperature
  less than 200\,K \citep{sternberg99}. Subsequently, the total column density and temperature
  are determined by fitting single- or two-temperature models to the
  excitation diagram. In a two-temperature fit, the first component
is used to constrain the column density and temperaure of warm gas and
the second component is for hotter gas. 

Since we only have two line detections (S(0) and S(1)) for regions 1,
2, 5, 6, 9, 11, 12, and three (S(0), S(1), S(2) or S(3)) for regions 3, 4
and 7 we apply a single-temperature fit to the observed column densities
(which is normalized to statistical weight $g_{\rm j}$, $N_{\rm j}/g_{\rm j}$) of
H$_2$ S(0) and S(1) transitions to constrain the total column density
and temperature. We performed a
least-square fit to determine the parameters, excitation temperature
$T_{\rm one}$ (from a single-temperature fit) and excited H$_2$ total column
density $N(\rm H_{2\,\rm ext}$). The single-temperature fits are shown in solid
black lines in the excitation diagrams, Fig. \ref{popdiag}, and the
derived parameters $T_{\rm one}$ and $N(\rm H_{2\,\rm ext}$) are given in
Table \ref{data3}. The excitation temperatures ($T_{\rm one}$) from the
single-temperature fits range from $86-137\,{\rm K}$. This temperature range corresponds to the ortho-to-para
ratio in the range $1.5-2.2$ (Fig. 1 in \citealt{sternberg99}), which is consistent with our initial assumption of 1.7.

In reality the ISM is made of gas with a distribution of temperatures.
It is very clear from the excitation diagram plot for regions 3, 4 and
7 in Fig. \ref{popdiag} that a single component fit does not pass
through the S(2) or S(3) points. This indicates that in general a
multiple-temperature fit is needed for characterizing the excitation
diagrams. It should also be noted that introducing a second
temperature in the fit may lower the excitation temprature of cooler
component relative to the single-temperature fit. Since the determination
of the total column density is very sensitive to the excitation
temperature, we might underestimate the total column density in a
single-temperature fit. In order to check these discrepancies we
performed a two-temperature model fit to regions 3, 4 and 7.
Although we have a measurement of third transition for these three
regions, in order to approximate a two-temperature fit we need
measurements for at least four transitions. Nevertheless, using these
three detections and upper limit measurements for higher transitions
we constrained the hot gas column densities and temperature by fitting
a two-temperature model for these three regions. The two-temperature
model fits in regions 3, 4 and 7 give the temperature for hot gas as
$<$750, $<$1100 and $<$630\,K respectively. These fits are shown as
dotted black lines in the excitation diagrams of regions 3, 4 and 7 in
Fig. \ref{popdiag} and the derived parameters $T$$_{\rm two,1}$ (low
temperature component from two-temperature fit), $T$$_{\rm two,2}$ (high
temperature component from two-temperature fit), are shown in
Table \ref{data3}. Note that the excitation temperature $T$$_{\rm
  two,1}$ derived from two-component fit is slightly lowered 
compared with single-temperature fit. As we have not detected any S(2) or S(3)
lines for regions 1, 2, 5, 6, 9, 11 and 12, and the two-temperature fit
through upper limit measurements only helps us to constrain an
upperlimit excitation temperature of a possible warmer component we use the column densities, $N(\rm H_{2\,\rm ext}$),
determined from single-temperature fit to estimate gas masses. Moreover,
the second component contributes a negligible fraction to the total
column density. The derived masses are given in Table. \ref{data3}. 
Note that we may underestimate
the mass if a second components is required, because inclusion of
second component can increase the final column densities. In order to
estimate this discrepancy, we compared the masses derived from single-temperature and
two-temperature fits for regions 3, 4 and 7. The two-temperature fit
increases the mass by a factor of 25\,per\,cent compared to the
single-temperature fit. 

\section{Comparison of warm H$_2$ with PAH and IR emission} 

\begin{figure}
\centering \epsfig{file=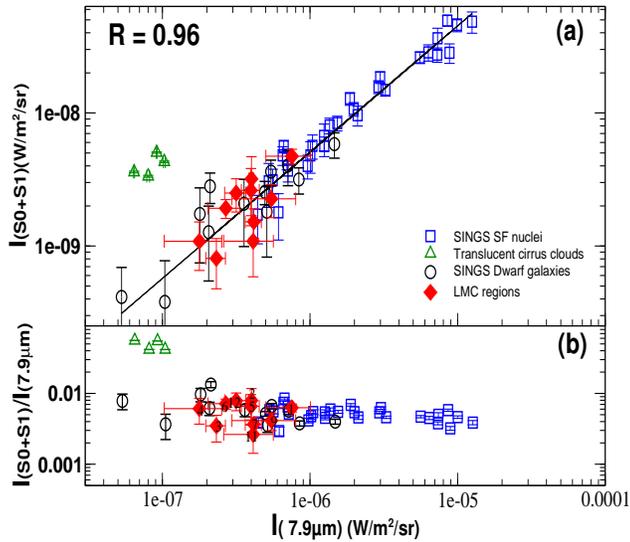, width=8.5cm, height=9.0cm,
  angle=-90}
  \caption{(a) The trend of power
    emitted in total S(0) and S(1) lines with the 7.9${\,\rm \mu m}$ PAH
    emission is fitted to a function log$(I_{\rm S(0)+S(1)}) = 0.95$ log$(I_{7.9\,\rm \mu m}) -
    3.0$ with a correlation coefficient 0.96. (b) This plot shows the ratio of power emitted in sum of H$_2$ lines S(0) and
    S(1) to power emitted in 7.9${\,\rm \mu m}$. Measurements for dwarf and star
    forming galaxy samples from SINGS data \citep{roussel07} and
    translucent cirrus clouds from \citet{ingalls11} are shown for
    comparison.}
  \label{PAH}  
\end{figure}

In order to quantify the importance of warm H$_2$ gas and thereby to
constrain possible excitation mechanism, we compare the power emitted
in H$_2$ with the dust emission traced in the form of PAH (traced by
IRAC 8.0${\,\rm \mu m}$ band emission), 24${\,\rm \mu m}$ and total infrared (TIR) surface
brightness. The TIR surface brightness is determined by combining the
IRAC 8.0${\,\rm \mu m}$, MIPS 24, 70${\,\rm \mu m}$ and 160${\,\rm \mu
  m}$ flux densities using equation\footnote{TIR$\sim$$0.95\braket{\nu
    I_\nu}_{7.9}+1.15\braket{\nu I_\nu}_{24}+\braket{\nu
    I_\nu}_{71}+\braket{\nu I_\nu}_{160}$} (22) in
\citet{draine07}. Perhaps, the scattered light from stellar emission could make
a significant contribution to the 8.0${\,\rm \mu m}$ fluxes. According
to J. Seok (private communication), the contribution of scattered
starlight in the 8.0${\,\rm \mu m}$ band in the H\,{\sc ii} region of the
LMC is $\sim$10\,per\,cent. Hence, we assume that the contribution of
starlight to the 8.0${\,\rm \mu m}$ band emission is less than $\sim$10\,per\,cent for
our sample regions.

In Fig. \ref{PAH} we examine the relation between power emitted in the
sum of H$_2$ S(0) and S(1) lines with the power emitted at IRAC 8.0${\,\rm
  \mu m}$ band, which traces the PAH 7.9$\,{\rm \mu m}$ emission in the ISM. For
comparison we show the SINGS galaxy samples \citep{roussel07} and
Galactic translucent clouds \citep{ingalls11} along with our ten LMC
regions. The ratio of the H$_2$ S(0) and S(1) lines to the
7.9$\,{\rm \mu m}$ PAH emission shows very little deviation for all
ten LMC regions, and agrees well with the SINGS normal galaxy
samples. In Fig. \ref{PAH}\,a we demonstrate that the relation between
H$_2$ and PAH emission can be fitted with the function log$(I_{\rm
  S(0)+S(1)}) = 0.95$ log$(I_{7.9\,\rm \mu m}) - 3.0$ with a
correlation coefficient of R$\sim$0.96, following the strong
correlation between H$_2$ and PAH emission in SINGS normal galaxies
earlier reported by \citet{roussel07}. In Fig. \ref{PAH} the LMC
samples are located along with the dwarf galaxies and these line-up
with the star-forming galaxies. If both H$_2$ and PAHs are
  excited in PDRs, such a tight correlation between H$_2$ and PAH
  emission is expected \citep{rigopoulou02}. In PDRs, the PAHs are
significant contributors to the photo-electric heating and H$_2$
emission is expected at the edge of the PDR. In the classic PDR of the
Orion Bar, it has been found that the H$_2$ emission is spatially
adjacent to the PAH emission \citep{tielens93}. In our observations,
we do not resolve the PDR layers; however, we expect the H$_2$ and PAH
emission to be correlated if the H$_2$ emission arises from a PDR. It
should also be noted here that, PDRs can be formed in any neutral
medium illuminated by FUV radiation where the radiation field is low,
compared to the intense radiation in star-forming regions. In such
regions, this kind of tight correlation between H$_2$ and PAH emission
is expected, even though the radiation field intensity is much lower
\citep{roussel07}.

\citet{ingalls11} have found excess H$_2$ emission compared to PAHs in
Galactic translucent clouds. They conclude that H$_2$ emission is dominated by 
mechanical heating even in non-active galaxies. This kind of excess H$_2$ emission has
also been reported in various other objects, such as active galactic
nuclei and ultraluminous infrared galaxies \citep{ogle10,
  higdon06}. They argue that an additional excitation
mechanism, possibly shock or gas heating by X-ray, might be responsible
for the excess emission of H$_2$. Such a heating mechanism can produce a
relatively larger enhancement of the warm H$_2$ compared to the UV
excitation in PDRs. In contrast, we argue that H$_2$ in the observed ISM of the LMC is dominated by FUV photo-processes.

\begin{figure}
\centering \epsfig{file=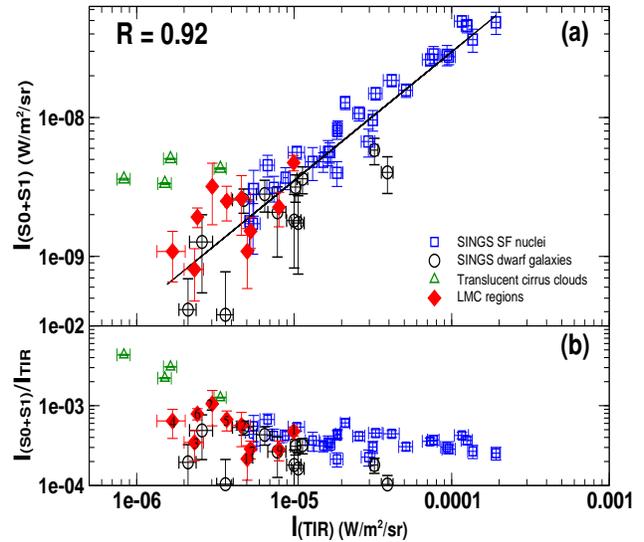, width=8.5cm, height=9.0cm, angle=-90}
  \caption{(a) The trend of power emitted
    in total S(0) and S(1) lines with the TIR surface brightness is
    fitted to a function log$(I_{\rm S(0)+S(1)}) = 0.93$ log$(I_{\rm TIR}) -
    3.82$ with a correlation coefficient 0.92. (b) This plot shows the ratio of power emitted in the sum of H$_2$ lines S(0)
    and S(1) to TIR surface brightness. Measurements for dwarf
    and star forming galaxy samples from SINGS data
    \citep{roussel07} and translucent cirrus clouds from
    \citet{ingalls11} are shown for comparison.}
  \label{TIR}  
\end{figure}

In Fig. \ref{TIR}\,b we plot the ratio of power emitted in the sum of
S(0) and S(1) lines to TIR power. \citet{roussel07} reported that the
H$_2$/TIR ratio is remarkably constant for their sample of normal
galaxies, and also in good agreement with the prediction from PDR
models by \citet{kaufman06}. Again, the LMC regions are shown
along with the SINGS dwarf galaxies and star-forming galaxies for
comparison. As in Fig. \ref{PAH}, the LMC sources have a similar trend
as the SINGS galaxies, but larger scatter. The ratio of H$_2$/TIR
power ranges over 0.0001 to 0.001. It is noted that the LMC points
are offset towards higher H$_2$/TIR ratios relative to the dwarf
galaxies, whereas they are consistent with the H$_2$/PAH ratios (Fig. \ref{PAH}). In Fig. \ref{TIR}\,a we show that the relation between
H$_2$ and TIR can be fitted to a funtion log$(I_{\rm S(0)+S(1)}) =
0.93$ log$(I_{\rm TIR}) - 3.82$ with a correlation coefficient
$\sim$0.92. In PDRs, the FUV radiation from young stars heats the dust
grains and give rise to infrared continuum radiation or mid-infrared
cooling lines such as H$_2$ rotational or fine structure lines; hence
we expect a strong tendency to increase in H$_2$ emission with
increase in total infrared flux. The ratio of fine structure line
[O{\,\sc i}] flux to TIR flux, [O{\,\sc i}]/TIR, has been used as a
diagnostic for the photo-electric heating efficiency in a variety of
environments of the LMC and SMC by \citet{vanloon10a, vanloon10b}.

\begin{figure}
\centering \epsfig{file=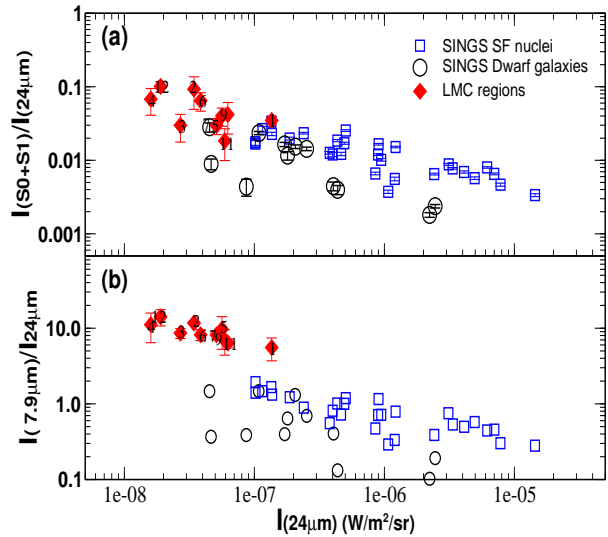, width=8.5cm, height=9.0cm, angle=-90}
  \caption{a) A trend of anti-correlation is noticed in the ratio
    H$_2$/24$\,{\rm \mu m}$ versus 24$\,{\rm \mu m}$ plot. The ratio of power emitted in the sum of H$_2$ S(0) and S(1) lines to
    power emitted in 24$\,{\rm \mu m}$ shows a large scatter among SINGS galaxy samples and
    LMC regions. b) This plot shows the ratio of power emitted in 7.9$\,{\rm \mu m}$ to 24$\,{\rm \mu m}$ as a function of 24$\,{\rm \mu m}$ power. The ratio tends to decrease with increase in 24$\,{\rm \mu m}$ emission.}
  \label{H2-24}  
\end{figure}

In order to investigate the relation of H$_2$ emission with the
24$\,{\rm \mu m}$ continuum emission, we plot the power ratio
H$_2$/24$\,{\rm \mu m}$ as a function of 24$\,{\rm \mu m}$ power in
Fig. \ref{H2-24}\,a. In star-forming galaxies, the H$_2$ luminosity
correlates with the 24$\,{\rm \mu m}$ emission that traces the amount of
star formation \citep{roussel07}. The H$_2$/24$\,{\rm \mu m}$ ratio
decreases with increase in 24$\,{\rm \mu m}$ emission, with a large
scatter, in Fig. \ref{H2-24}\,a. Note that here we compare the H$_2$
and 24$\,{\rm \mu m}$ emission of ten different regions of the LMC
with those in SINGS galaxies. This anti-correlation may indicate
that globally, the 24$\,{\rm \mu m}$ emission traces a significantly
different, mostly uncorrelated medium from the warm H$_2$ gas, in
contrast to the PAH 7.9$\,{\rm \mu m}$ emission (see
Fig.~\ref{PAH}). In order to better understand this relation, in
Fig. \ref{H2-24}\,b we plot the power ratio 7.9$\,{\rm \mu
  m}$/24$\,{\rm \mu m}$ as a function of 24$\,{\rm \mu m}$ power. This
plot clearly shows an anti-correlation dependency of PAH emission with
24$\,{\rm \mu m}$ emission. This trend could be due to the increased starlight 
intensities in regions where the 24$\,{\rm \mu m}$ emission is higher \citep{draine07}, or even 
combined with the preferential destruction of small PAHs in these regions \citep{draine01}.
 If the starlight intensity $U$ is sufficiently high, the dust
  (i.e., nano dust) which emits at 24$\,{\rm \mu m}$ is no longer in
  the regime of single-photon heating. Therefore $I_{24\,\rm \mu m}/\it U$ increases with $U$, while the 7.9$\,{\rm \mu m}$ emitter
  (PAHs) is still in the regime of single-photon heating. Hence
  $I_{7.9\,\rm \mu m}/\it U$ does not vary with $U$ (see Fig. 13\,b of
  \citealt{draine07}), and $\frac{I_{7.9\,\rm \mu m}}{I_{24\,\rm \mu
    m}} = \frac{{I_{7.9\,\rm \mu m}/\it U}}{{I_{24\rm\,\mu m}/\it U}}$
  intends to decrease with $I_{24\,\rm \mu m}$ and $U$. Furthermore,
  we cannot rule out the possibility that the 7.9$\,{\rm \mu m}$
  emitter may be partly destroyed at higher $U$, which would also lead
  to a decrease of power ratio 7.9$\,{\rm \mu m}$/24$\,{\rm \mu m}$
  with 24$\,{\rm \mu m}$ power. If PAHs are partly destroyed one would
  expect a decreased photo-electric heating of H$_2$ and thus a
  decrease of power ratio H$_2$/24$\,{\rm \mu m}$ with 24$\,{\rm \mu
    m}$ power and $U$.

\section{Warm H$_2$, cold H$_2$, H{\,\sc i} and H$\alpha$: phases of the ISM}

We can compare the amount of gas contained in the warm component
with the major phases of hydrogen: cold molecular H$_2$, atomic H{\,\sc i}, and
ionized H$^+$. The
cold H$_2$ gas mass is derived using CO (J=1-0) line
observations \citep{wong11}
obtained with the Magellanic Mopra Assessment
(MAGMA) and the NANTEN surveys \citep{fukui08}. As part of MAGMA project the LMC
has been surveyed in CO (J=1-0) line with the 22\,m Mopra telescope of
the Australian Telescope National Facility \citep{wong11}. This
observation achieved a 45$\,$arcsec angular resolution with a
sensitivity nearly 0.4$\,{\rm K}\,{\rm km\,s}^{-1}$. At this angular resolution, the Mopra
CO intensity map is sufficient for our analysis. However, the MAGMA
survey does not cover the whole LMC, as it was a targeted follow-up
survey of the LMC regions where significant CO emission is detected in the 
previous NANTEN CO(J=1-0) survey with angular resolution
2.6$\,$arcmin. Hence, we have to use the NANTEN CO intensities for at
least five regions where no MAGMA observations exist. The CO integrated intensities were extracted over a
30$\,$arcsec radius aperture size. These integrated intensities (${\rm K}\,{\rm km\,s}^{-1}$) were then converted to H$_2$
(cold) column densities assuming the Galactic CO-to-H$_2$ conversion
factor of 2$\times$10$^{20}\,{\rm cm}^{-2}\,(\rm {K\,km\,s}^{-1})^{-1}$ \citep{bolatto13}. 

The atomic gas is traced by $21\,{\rm cm}$ line
emission observations by the Australia Telescope Compact Array (ATCA)
and Parkes single dish telescope spanning 11.1$\times$
12.4\,deg$^2$ on the sky at a spatial resolution 1$\,$arcmin
\citep{kim03}. The construction of a H{\,\sc i} intensity map from the H{\,\sc i} data
cube and further conversion to column density is discussed by
\citet{jean08}. We integrated the H{\,\sc i} column densities within an
aperture radius 30$\,$arcsec. The derived H{\,\sc i} column densities and masses are given in Table
\ref{data4}. 

The ionized gas is traced by the H$\alpha$ emission observed in the
Southern H-Alpha Sky Survey Atlas (SHASSA; \cite{gaustad01}) carried
out at the Cerro Tololo Inter-American Observatory in Chile. The
angular resolution of that image is 0.8$\,$arcmin with a sensitivity
level of 2\,Rayleighs, where 1 Rayleigh = $2.25\,{\rm pc}\,{\rm
  cm}^{-6}$ for T$_{\rm e}$=$8000\,{\rm K}$ \citep{dickinson03},
 corresponding to $2.41\times10^{-7}$\,erg\,s$^{-1}$cm$^{-2}$\,sr$^{-1}$. In
order to determine the H$^+$ column density, we first calculated the
electron density $n_{\rm e}$ using the formula EM=$n_{\rm e}^{2}$L
\citep{dickinson03}, where EM is the emission measure and L is the
path length in parsec. We assume here the width of emission feature is
the same as the emitting path length along the line of
sight. Consequently, the H$^+$ column density is calculated from the H$\alpha$ intensity and electron density of each region using
equation\footnote{{\begin{equation} \frac{N(H^{+})}{H\,cm^{-2}}=1.37
      \times 10^{18}\frac{I_{H\alpha}}{R}(\frac{n_{\rm e}}{{\rm
          cm}^{-3}})^{-1} \end{equation}}} (6) reported by
\citet{jean08}.

\begin{figure*}
\centering 
\includegraphics[trim=0.0cm 1.3cm 0.0cm 1.5cm, clip = true, scale=0.6]{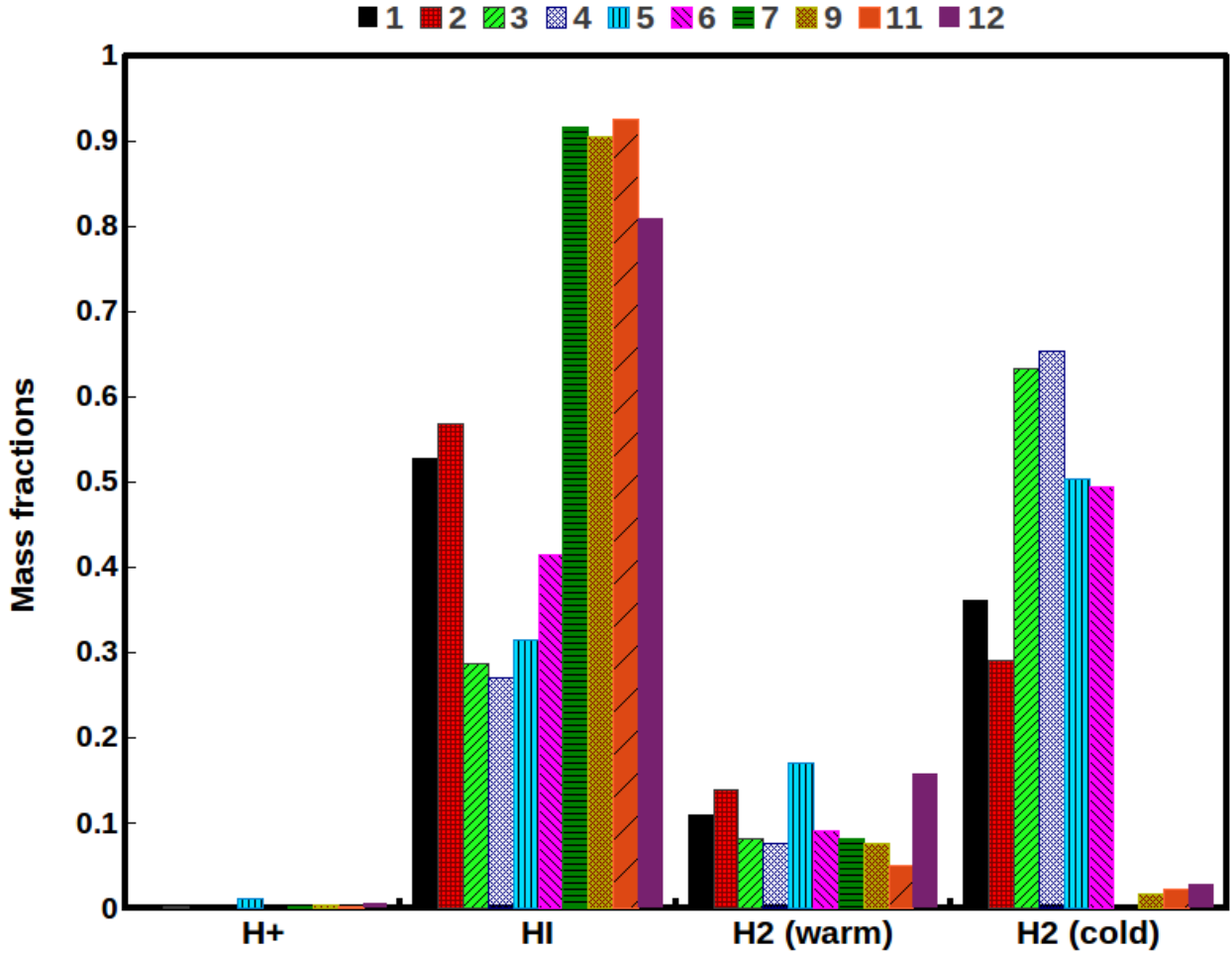}
  \caption{A histogram showing mass fractions of H$^{+}$, H{\,\sc i}, warm
    H$_2$ and cold H$_2$ in regions where the H$_2$ excitation is
    detected. The colors indicate ten different regions. }

  \label{his}  
\end{figure*}

Our main purpose here is to estimate what fraction of warm H$_2$ takes
up the total gas, compared to the other phases of the ISM represented
by cold H$_2$, H{\,\sc i} and H$^+$. Mass fractions in atomic, ionic,
warm H$_2$ and cold H$_2$ form are given in Table \ref{data4} and a
histogram is shown in Fig. \ref{his}. 
The warm H$_2$ mass fraction varies slightly from region to region with a
significant contribution $\sim5-17$\,per\,cent to the total gas mass. The
histogram in Fig. \ref{his} compares the fraction of H{\,\sc i}, warm
H$_2$ and cold H$_2$ in ten regions where H$_2$ excitation is
detected. Region 5 shows the largest amount of warm H$_2$, 17\,per\,cent of
the total gas mass (see Table \ref{data4}), with atomic gas and cold
H$_2$ masses at nearly 31\,per\,cent and 50\,per\,cent respectively. Warm H$_2$ is
$5-15$\,per\,cent of the total gas mass in regions 1, 2, 7, 9 and 12. The
H$_2$ traced by NANTEN CO is very negligible and the atomic gas is
$\ge80$\,per\,cent of the total gas (see Table \ref{data4}) in regions 7, 9, 11
and 12. In particular, regions 1, 2, 7, 9, 11 and 12 are diffuse
atomic, where H{\,\sc i} is the dominant ISM component, while regions 3, 4,
5 and 6 are molecular dominated. The amount of warm H$_2$ is equally
important in both atomic and molecular dominated regions. In most
regions only two lines are detected, hence the mass of warm H$_2$
could be underestimated. In fact, a large uncertainty is expected in
the determination of cold H$_2$ mass from CO. The first dominant
uncertainty comes from matching the size of the region over which the IRS spectrum was integrated with two different
beam sizes for CO observations obtained from Mopra and NANTEN
telescopes. Since, the CO luminosity for regions 2, 9, 11 and 12 were
extracted from NANTEN CO map with angular resolution 2.6$\,$arcmin, we
expect a large uncertainty in the cold H$_2$ mass for those
regions. Another source of uncertainty comes from the CO-to-H$_2$
conversion factor for which we used a Galactic value of
2$\times$10$^{20}\,{\rm cm}^{-2}\,({\rm K}\,{\rm km}\,{\rm
  s}^{-1})^{-1}$ according to \citet{bolatto13}. Note that the
CO-to-H$_2$ conversion factor strongly depends on metallicity which we
assumed uniform over the regions and the applicability of the Galactic
value is debated.

\begin{figure*}

\includegraphics[trim=2.8cm 0cm 0cm 2cm, clip=true, scale=0.25, angle=-90]{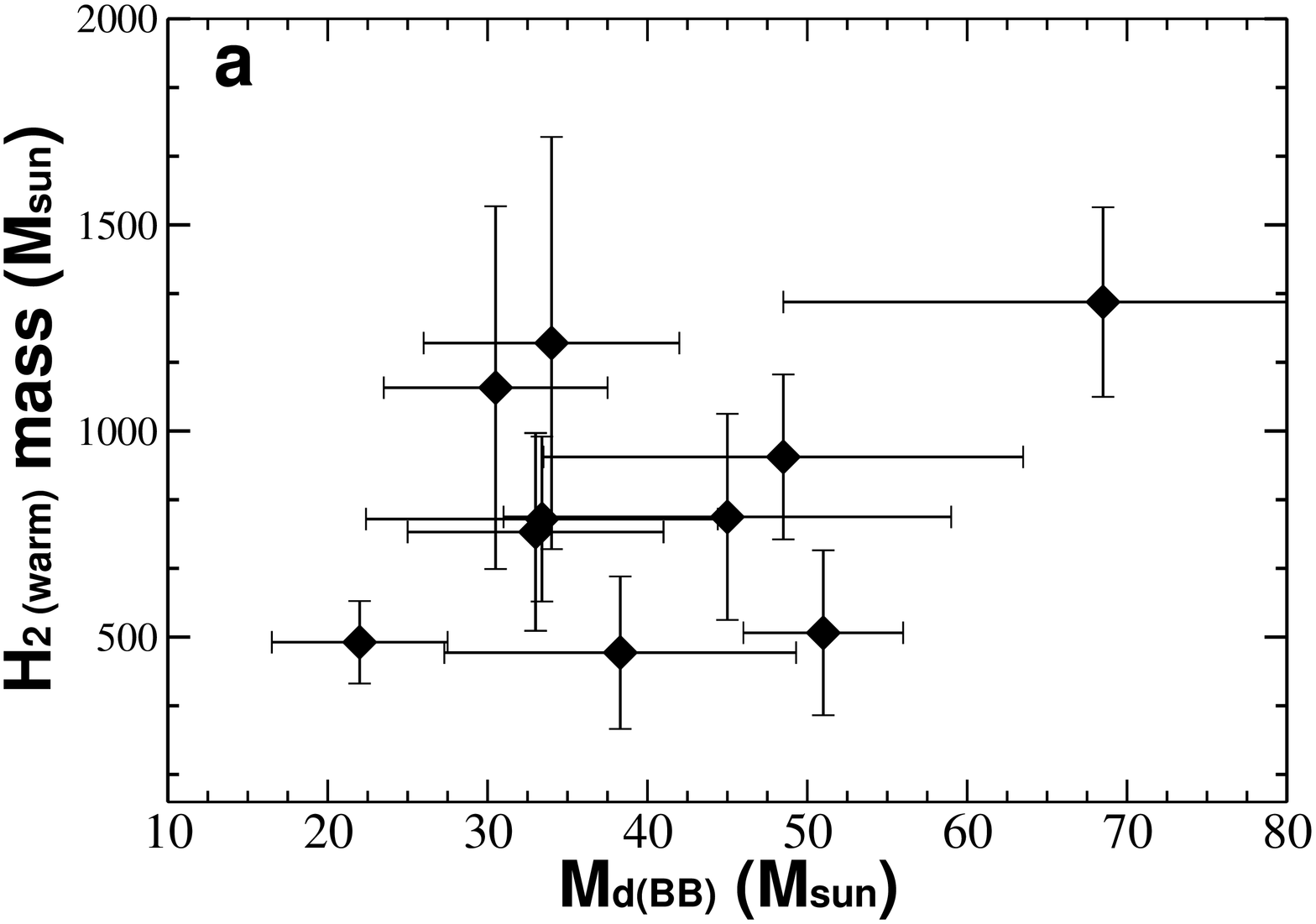}
\includegraphics[trim=2.8cm 0cm 0cm 2cm, clip=true, scale=0.25, angle=-90]{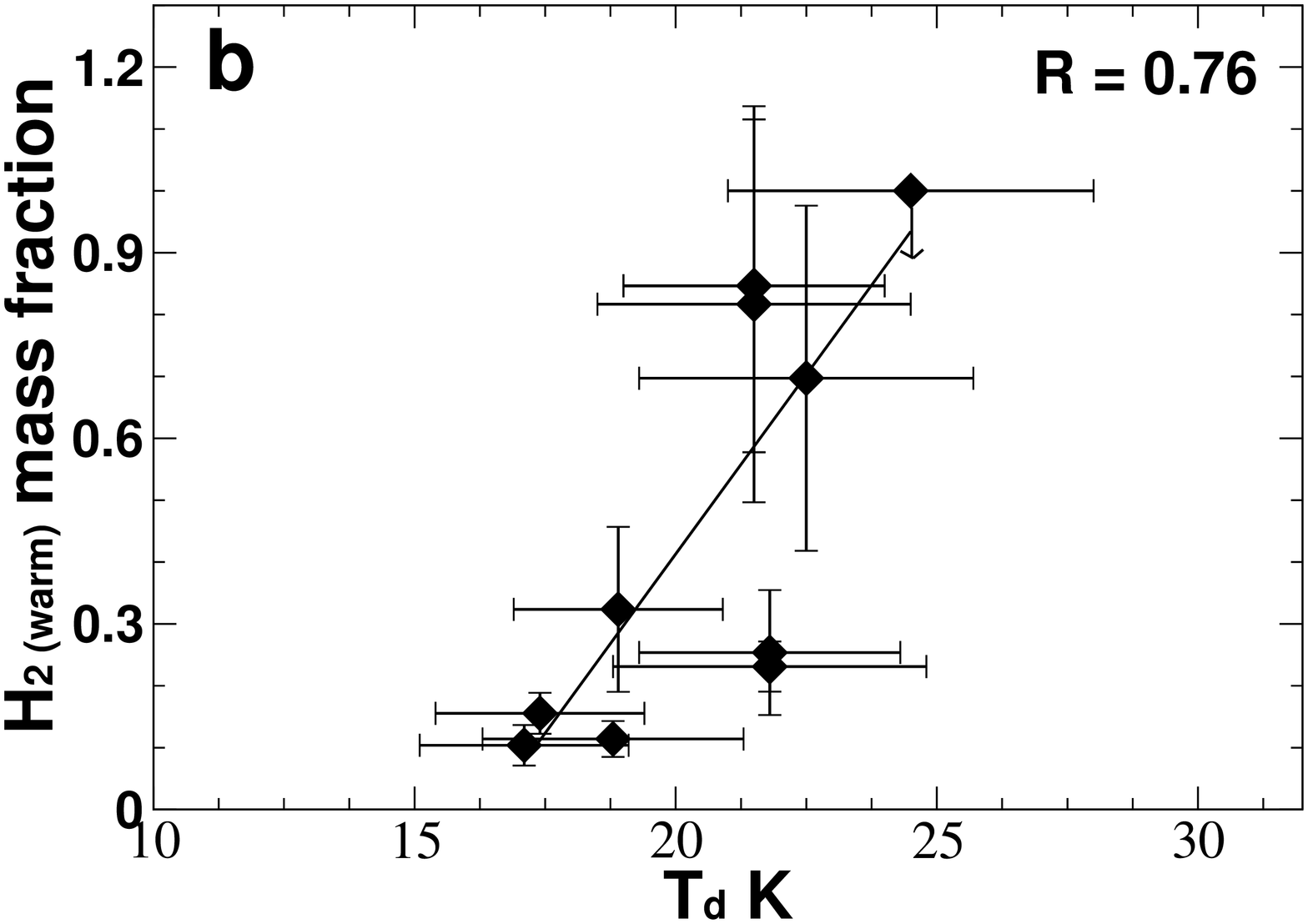}\\
\includegraphics[trim=2.8cm 0cm 0cm 2cm, clip=true, scale=0.25, angle=-90]{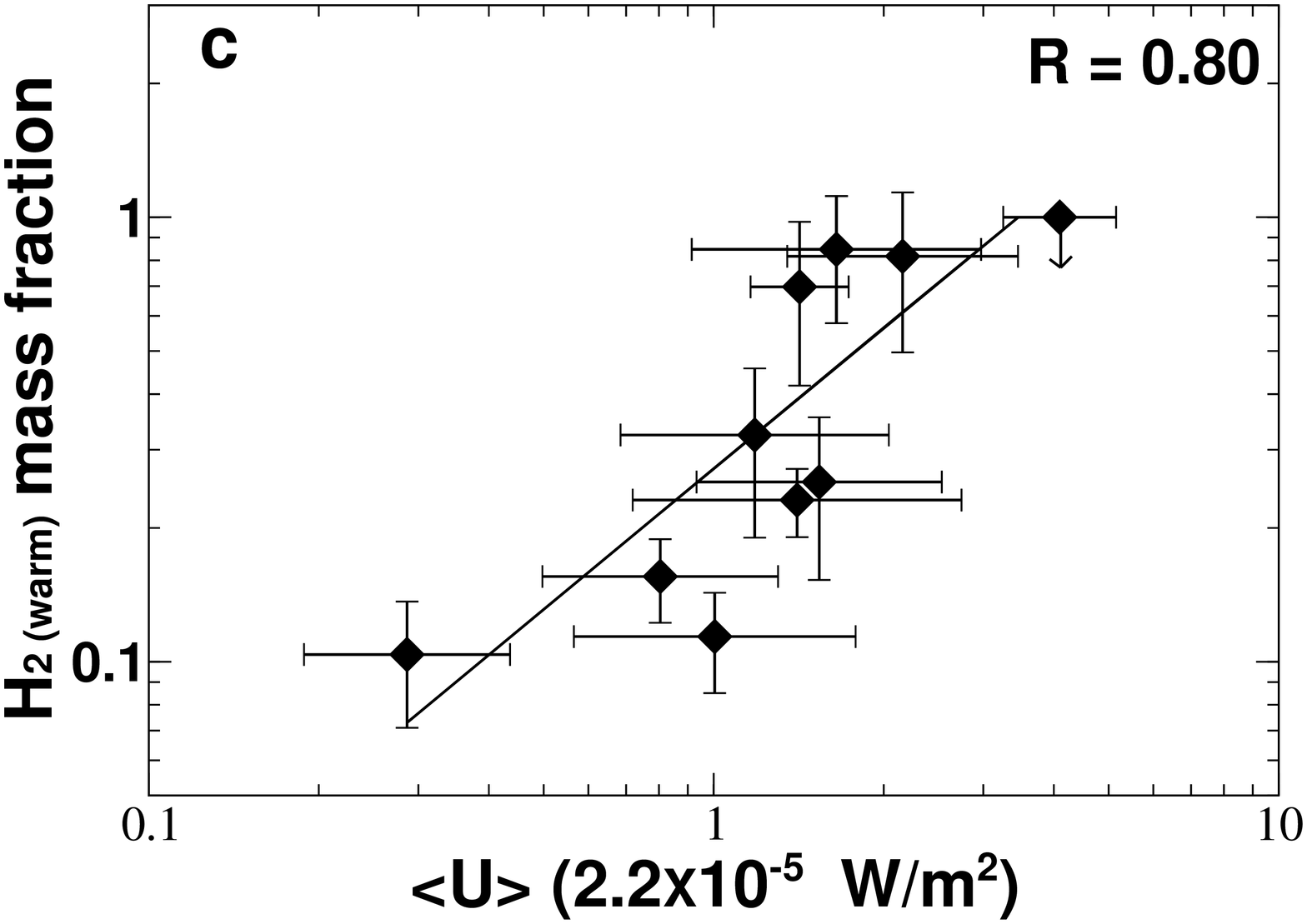}
\includegraphics[trim=2.8cm 0cm 0cm 2cm, clip=true, scale=0.25, angle=-90]{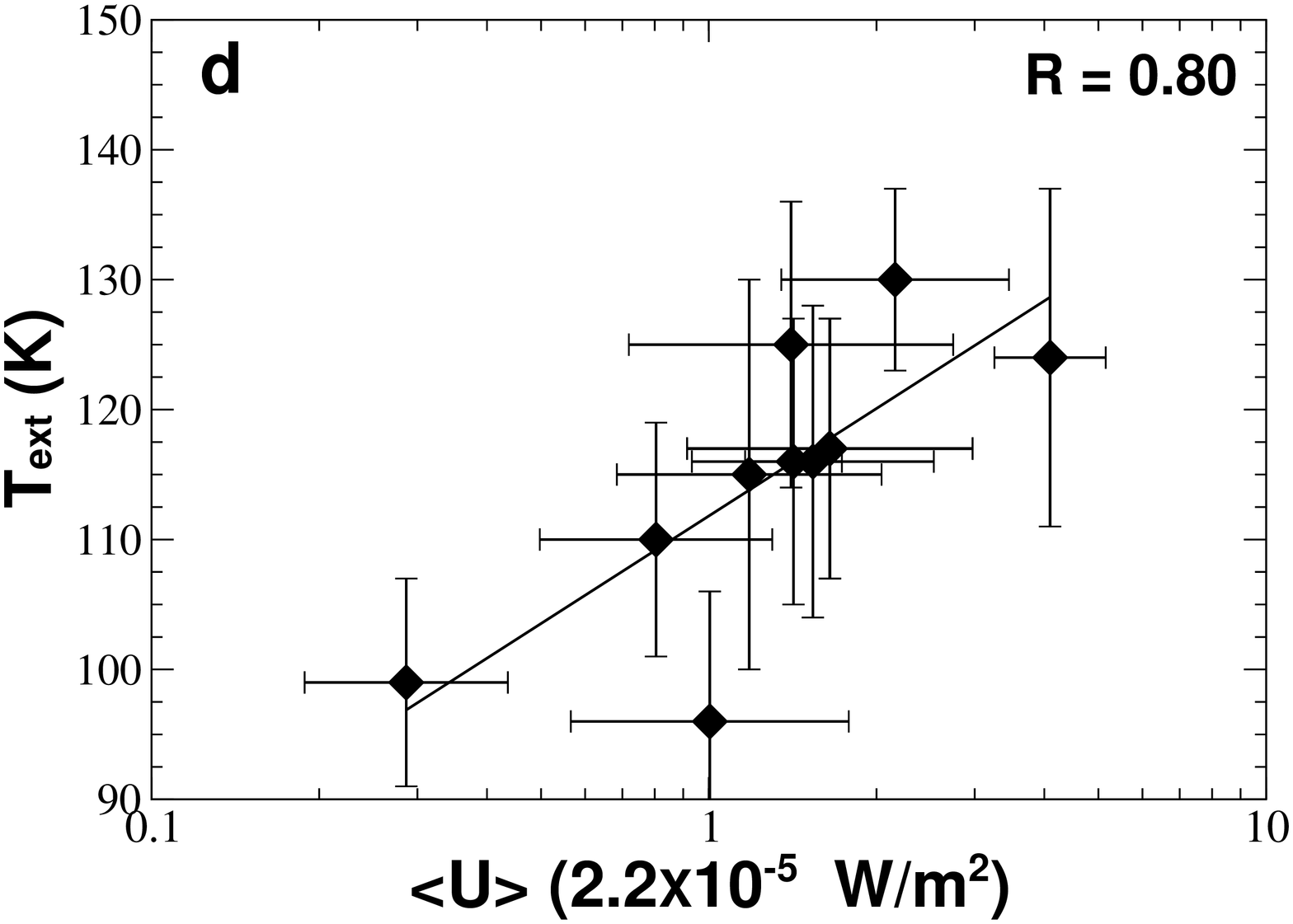}\\
\includegraphics[trim=2.8cm 0cm 0cm 2cm, clip=true, scale=0.25, angle=-90]{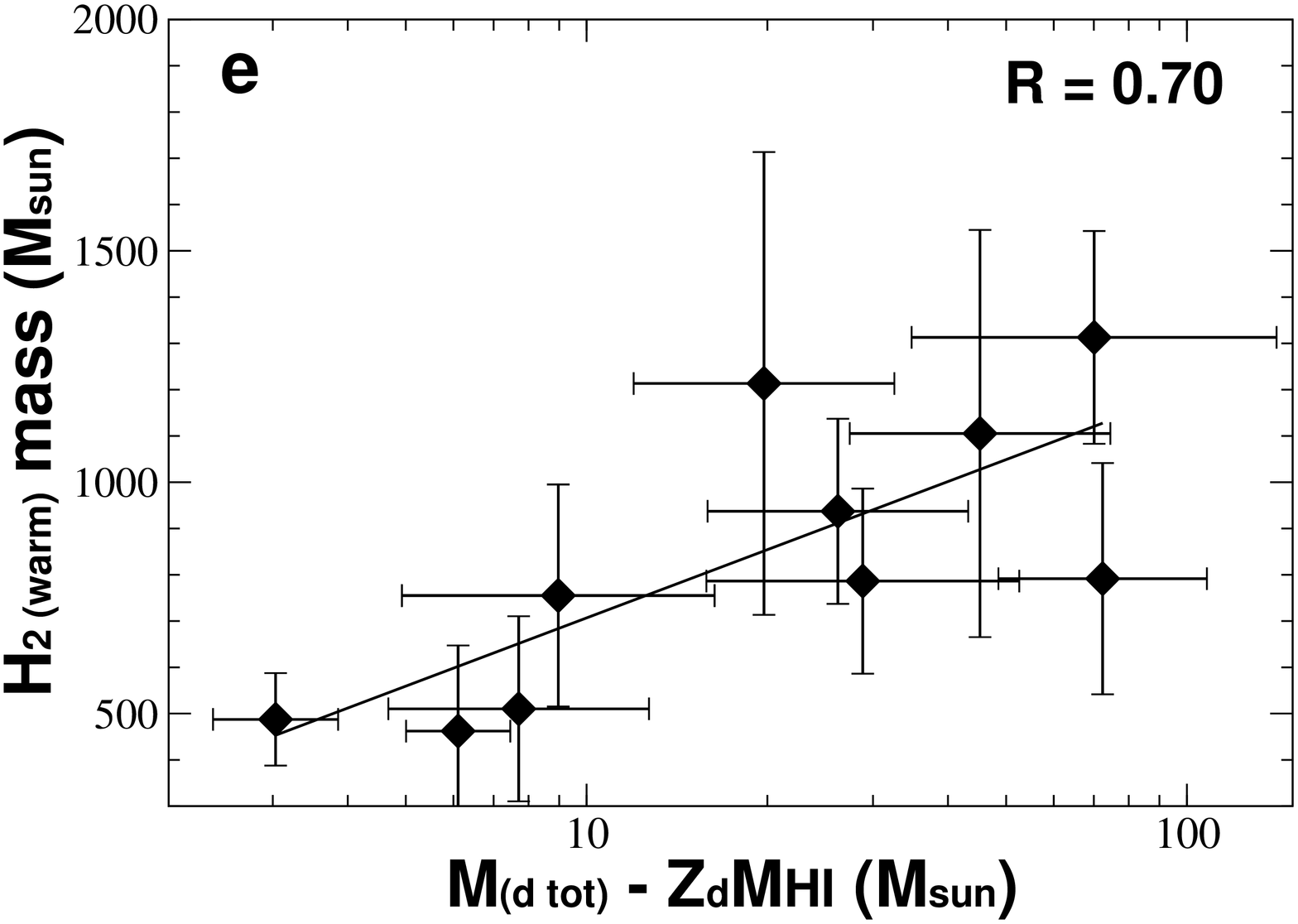}

\caption{The correlation tests for warm H$_2$ with the dust parameters. The
correlation coefficients (R) are given on those plots which show a
noticeable trend. a) The warm H$_2$ mass versus dust mass M$_{\rm d\,(BB)}$, derived from modified blackbody fit. b) The warm H$_2$ mass fraction versus dust equilibrium temperature. c) The H$_2$ warm mass
fraction as a function of average starlight intensity $\braket{U}$. d) The H$_2$ excitation
temperature T$_{\rm ext}$ versus the average
starlight intensity $\braket{U}$. e) The H$_2$ warm mass
is plotted against the total dust mass derived by fitting the IR SED
with the \citet{galliano11} model. The dust associated with H{\,\sc
  i} is removed by a factor Z$_{\rm d}$M(H{\,\sc i}), where
Z$_{\rm d}$ is dust-to-gas ratio and M(H{\,\sc i}) is total H{\,\sc i}
gas mass.}

  \label{corr1} 
\end{figure*}

\section{Comparison of warm H$_2$ with dust}

An important ISM component that is closely linked to the formation of
H$_2$ is dust. Hence, for a first approximation here we try to
determine the physical parameters of dust by fitting modified blackbody
curves to the far-infrared spectral energy distributions (SEDs). We determine the temperature and mass of
the dust by fitting a single-temperature modified blackbody
curve to the far-infrared SED obtained from {\it Herschel} photometric
observations. PACS (100 and 160\,${\rm \mu m}$) and SPIRE (250, 350
and 500\,${\rm \mu m}$) fluxes are fitted with the modified blackbody
curve, following the method outlined in \citet{gordon10}. We restrict
our SED fitting above 70\,${\rm \mu m}$ as we assume the dust grains
are in thermal equilibrium with the radiation field.
In our fitting method, the emissivity $\beta$ is allowed to vary
between 1\,$<$\,$\beta$\,$<$\,2.5 \citep{gordon14}. The uncertainties
in mass and temperature come from SED fitting, where quadratic sum of
uncertainties in flux extraction from photometric images and absolute
error in flux calibration are adopted \citep{meixner13}. The fluxes are
randomly perturbed within a Gaussian distribution of uncertainties
and chi-square fits are used to determine dust mass and
temperature. The measurements are given in Table. \ref{data4}. The
dust shows a range in mass 15$\le$M$_{\rm
  d(BB)}\le$90$\,$M$_{\odot}$ within the 1\,arcmin integrated region
and in temperature 15$\le$$T_{\rm d}\le$28$\,{\rm K}$.  

In the far-infrared domain, the emission is dominated by cold large
grains which are in thermal equilibrium with the radiation
field. Therefore the spectrum can be approximated with a modified
blackbody curve. This is not well constrained for mid-infrared continuum,
where the emission is dominated by small grains and PAHs. 
Moreover, the modified blackbody fits are known to
underestimate the dust by a factor of nearly 0.7 in most regions of
the LMC \citep{galliano11}. Hence, in order to constrain the total
dust mass with better uncertainties, we fitted the SEDs in the
wavelength range from 8.0 to 500${\,\rm \mu m}$ (combining the IRAC,
MIPS and {\it Herschel} SPIRE bands) with the \citet{galliano11} model
using a least-square approach and Monte-carlo error propagation. More
details of fitting and dust parameter determination can be found in
\citet{galliano11}. The derived total dust masses M$_{\rm d\,tot}$ are
given in Table \ref{data4} for comparison with the dust mass derived from
single-temperature blackbody fits. It should be noted that, error bars
on dust mass derived from this method are highly asymmetric (Table
\ref{data4}). In addition, we derive the average starlight intensity
$\Braket{U}$ using equation (9) of \citet{galliano11}: this parameter
is on first approximation related to the grain equilibrium temperature
by $T_{\rm d}\simeq U^{1/5.7}\times19\,{\rm K}$.

In Fig. \ref{corr1} we examine the relation of warm H$_2$ with the
dust parameters derived by different methods explained above. No clear
relation is noticed for the warm H$_2$ mass with dust mass
derived by fitting a blackbody curve to far-infrared SED
(Fig. \ref{corr1}\,a). The warm H$_2$ mass fraction (in terms of the
total H$_2$ mass) is found to be positively correlated (R$\sim$0.76)
with the dust equilibrium temperature (Fig. \ref{corr1}\,b).
Fig. \ref{corr1}\,c and d show that the H$_2$ mass fraction and excitation
temperature are positively correlated (R$\sim\,$0.80) with the average
starlight intensity $\Braket{U}$, which is a measure of dominant input
energy for dust heating in the medium. These correlation tests confirm
that photo-electric heating in the PDR is efficient to excite H$_2$ in
both atomic and molecular dominated diffuse regions. In
Fig. \ref{corr1}\,e, the warm H$_2$ mass shows a moderate positive
correlation (R$\sim$0.70) with the total dust mass, derived by fitting
infrared SED with the \citet{galliano11} model. Since we compare dust
mass with the H$_2$ mass, we attempt to remove the dust associated
 with H{\,\sc i} from the total dust mass. Hence, we subtract a
factor Z$_{\rm d}$M(H{\,\sc i}) from dust mass where Z$_{\rm d}$ is
dust-to-gas ratio and M(H{\,\sc i}) is total H{\,\sc i} gas mass.

\begin{table*}
\caption{Observed H$_2$ and dust surface brightness in units of $\rm W\,m^{-2}\,sr^{-1}$.}
\label{lines}
\begin{tabular}{cccccccccccc}
\hline
\noalign{\smallskip}
Region  & S (0) & S(1) & S(2) & S(3) & S(4) & S(5) & S(6) & S(7)& 7.9$\,{\rm \mu m}$ & 24$\,{\rm \mu m}$ & TIR \\
        & 10$^{-09}$ & 10$^{-09}$ & 10$^{-09}$ &10$^{-09}$  & 10$^{-09}$ & 10$^{-09}$ & 10$^{-09}$ &10$^{-09}$&10$^{-07}$ &10$^{-08}$ &10$^{-06}$\\
\hline
 1  & 2.2(0.3)  & 2.5(0.3)&$<1.8$    & $<2.8$    & $<2.2$ & $<1.4$ &$<2.3$ &$<5.6$ & 7.54(2.5)  &13.6(0.01)&9.92(0.2)\\
 2  & 1.9(0.8) & 1.3(0.7) & $<2.8$  & $<3.8    $ & $<0.6$ & $<1.1$ &$<9.1$ & $<2.0$& 4.01(0.2)  &3.43(0.01)&3.02(0.06)\\
 3  & 0.62(0.2) & 0.18(0.1) & 0.15(0.03) & $<2.0$  & $<1.7$ & $<1.9$ & $<4.4$& $<6.1$&2.32(0.4) &2.70(0.005)&2.33(0.3) \\
 4  & 0.80(0.4) & 0.30(0.03)& 0.60(0.5)& $<1.6$    & $ <3.0$& $<2.9$ &$<2.3$ & $<4.7$&1.78(0.8)  &1.60(0.005)&1.60(0.3)\\
5   & 1.3(0.7)  & 1.3(0.5) & $<0.7$    & $<0.81$    & $<0.93$ &$<1.2$  &$<2.9$ &$<3.2$ & 3.98(0.6) &6.25(0.002)&4.63(0.6) \\
6   & 1.2(0.2)   & 0.72(0.2) & $<0.8$    & $< 2.2$& $<2.4$    & $<1.7$ & $<3.3$& $<4.9$&2.69(0.7)  &1.90(0.006)&2.43(0.03)\\
7   & 0.62(0.1)   & 0.9(0.3)& $<0.8$    & 1.4(0.08) & $<1.6$ & $<2.6$ & $<2.5$& $<2.5$&4.16(0.5)  &5.11(0.01)&5.29(0.06)\\
9   & 0.73(0.3)  & 1.5(0.3)& $<3.9$    & $<6.5$    & $<4.0$ & $<9.0$ & $<9.4$& $<9.7$&5.48(2.5)  &5.64((0.007)&8.01(0.08)\\
11  & 0.51(0.2)   & 0.58(0.3) & $<1.2$    & $<3.5$    &$<3.4$  & $<7.1$ & $<5.2$& $<9.1$&4.12(1.5)  &5.92(0.01)&5.03(0.06)\\
12  & 1.2(0.3)   & 1.3(0.4)& $<1.3$    &  $<2.6$   & $<2.7$ & $<8.3$ &$<6.0$ & $<9.5$&3.16(0.5)  &3.86(0.006)&3.70(0.1)\\

\hline
\end{tabular}
\end{table*}

\begin{table*}
\caption{Observed H$_2$ column densities measured from line intensities.}
\label{data2}
\begin{tabular}{ccccccccc}
\hline
\noalign{\smallskip}
Region  & S (0) & S(1) & S(2) & S(3) & S(4) & S(5) & S(6) & S(7) \\
        & 10$^{19}$ & 10$^{17}$ & 10$^{16}$ &10$^{16}$  & 10$^{15}$ & 10$^{15}$ & 10$^{15}$ &10$^{15}$  \\
        & $\rm cm^{-2}$ & $\rm cm^{-2}$ & $\rm cm^{-2}$ &$\rm cm^{-2}$  & $\rm cm^{-2}$ & $\rm cm^{-2}$ & $\rm cm^{-2}$ &$\rm cm^{-2}$  \\
\hline
 1  & 1.4(0.15)  & 6.7(0.5)&$<5.1$    & $<1.7$    & $<4.3$ & $<1.1$ &$<0.8$ &$<0.98$ \\
 2  & 1.2(0.5) & 3.1(1.7) & $<8.0$  & $<2.3    $ & $<1.2$ & $<0.82$ &$<3.0$ & $<0.35$\\
 3  & 0.38(0.09) & 0.42(0.3) & 0.42(0.07) & $<1.3$  & $<3.3$ & $<1.4$ & $<1.5$& $<1.0$\\
 4  & 0.47(0.22) & 0.67(0.06)& 1.6(1.4)& $<1.0$    & $ <5.4$& $<2.2$ &$<0.8$ & $<0.8$\\
5   & 0.80(0.4)  & 3.0(1.1) & $<1.9$    & $<0.51$    & $<1.8$ &$<0.9$  &$<1.0$ &$<0.57$ \\
6   & 0.96(0.13)   & 0.92(0.2) & $<2.2$    & $< 1.2$& $<4.5$    & $<1.5$ & $<1.1$& $<0.49$\\
7   & 0.38(0.08)   & 2.2(0.6)& $<2.4$    & 0.87(0.05) & $<3.0$ & $<1.9$ & $<0.8$& $<0.43$\\
9   & 0.45(0.2)  & 3.5(0.7)& $<10.0$    & $<4.0$    & $<7.7$ & $<7.6$ & $<4.6$& $<3.0$\\
11  & 0.31(0.13)   & 1.3(0.6) & $<3.4$    & $<2.2$    &$<6.6$  & $<5.3$ & $<1.8$& $<2.0$\\
12  & 0.78(0.19)   & 2.9(0.98)& $<3.8$    &  $<1.6$   & $<5.2$ & $<6.2$ &$<2.1$ & $<2.5$\\

\hline
\end{tabular}
\end{table*}

\begin{table*}
\caption{Excitation temperature, total column densities and mass of
  excited H$_2$ derived from excitation diagrams. For regions 3, 4 and
  7 the excitation temperatures for both single-temperature and
  two-temperature fits are presented.}
\label{data3}
\begin{tabular}{l l l l l l l l l l l l l}
\hline
\noalign{\smallskip}
Reg  & \multicolumn{3}{c}{$T_{\rm ext}$}&$N(\rm H_{2\,\rm ext}) $&M(H$_{2\,\rm ext})\,$  \\
 & $T_{\rm one}$ &$T_{\rm two,1}$ &$T_{\rm two,2}$ &${\rm cm}^{-2}$&M$_{\odot}$   \\
 & K& K& K&10$^{20}$&10$^{3}$      \\

\hline
 1  & $126\pm11$  &  &         & $4.9\pm0.9$  &$1.31\pm0.23$        \\
 2  & $115\pm15$   &  &        & $4.6\pm1.9$  &$1.21\pm0.50$        \\
 3  & $96\pm10$      &$90$\tiny{$^{+10}_{-5}$}  & $<750$& $3.0\pm0.7$&$0.79\pm0.20$    \\
 4  & $99\pm8$     &$87$\tiny{$^{+15}_{-7}$}  & $<1100$ & $3.0\pm0.9$&$0.79\pm0.25$     \\
 5  & $116\pm12$     &      &       & $4.2\pm0.3$  &$1.1\pm0.45$    \\
 6  & $110\pm9$      &      &       & $3.5\pm0.8$    &$0.94\pm0.20$ \\
 7  & $124\pm13$     &$100$\tiny{$^{+7}_{-7}$}  & $<630$ & $1.8\pm0.4$  &$0.49 \pm0.10$\\
  
 9  & $130\pm7$     &       &        & $1.9\pm0.8$      &$0.51 \pm0.20$   \\ 
 11 & $116 \pm11$       &     &      & $1.7\pm0.7$  & $0.46 \pm0.20$ \\
 12 & $117\pm10$     &       &      & $2.9\pm0.9$  & $0.75\pm0.25$      \\ 
\hline
\end{tabular}

\end{table*}

\begin{table*}
\caption{Dust and gas physical parameters, mass fractions in atomic, ionized, warm molecular and cold molecular phases of ten regions where the H$_2$ excitation is detected.}
\label{data4}
\begin{tabular}{l l l l l l l l l l l l l l l}
\hline
\noalign{\smallskip}
Reg  & $T_{\rm d}$&$^1$M$_{\rm d\,(BB)}$  & $^2$M$_{\rm d\,tot}$  &   $N$(H{\,\sc i}) & M(H{\,\sc i})&$^3$$N(\rm H_{\rm 2\,cold})$ &M(H$_{\rm 2\,cold})$&$^4$M$_{\rm tot}$ & f(H{\,\sc i}) & f(H$^{+}$) & f(H$_{\rm 2\,warm})$ & f(H$_{\rm 2\,cold}$) \\
     &   &  &  &        10$^{21}$ &10$^{3}$ &10$^{21}$ &10$^{3}$ &10$^{3}$ & & & &\\
 &K&M$_{\odot}$&M$_{\odot}$  &${\rm cm}^{-2}$ &M$_{\odot}$ &${\rm cm}^{-2}$ &M$_{\odot}$ &M$_{\odot}$& & & &\\
\hline
 1  & $21.8\pm 3.0$   &$68.5\pm20$ & 148\tiny{$^{ +150}_{-74}$}  &4.8   & 6.4   & 1.6       & $4.5 $       &12.0  & 0.53   & 0.004     & 0.11     & 0.36    \\
 2  & $18.9 \pm 2.0$    &$34.0\pm8.0$  & 40\tiny{$^{+26}_{-15}$}   & 3.7   & 5.0  & $^{\tiny{5}\,}0.09$       & $ 2.5 $      &8.7  & 0.57   & 0.002     & 0.14     & 0.29       \\
 3  & $18.8 \pm 2.5$  &$33.4\pm11$ & 40\tiny{$^{+33}_{-18}$}  &  2.1  & 2.8 & 2.3      & $6.1$            &9.7   & 0.30  &    $--$    & 0.08     & 0.63      \\
 4  & $17.1\pm2.0$    &$45.0\pm14$ & 99\tiny{$^{+48}_{-32}$}   & 2.1   & 2.8 & 2.6       & $6.8$          &10.5   & 0.27  &   $--$     & 0.08     & 0.65       \\
 5  & $21.8 \pm 2.5$ &$30.5\pm7.0$   & 66\tiny{$^{+43}_{-26}$}    & 1.5    & 2.0 & 1.2        & $3.2$       &6.5    & 0.31  & 0.01      & 0.17     & 0.50     \\
 6  & $17.4\pm2.0$   &$48.5\pm15$  & 44\tiny{$^{+29}_{-18}$}    & 3.2   & 4.3 & 1.9       & $5.1$         &10.3   & 0.41  & $--$       & 0.09     & 0.50   \\
 7  & $24.5\pm3.5$   &$22.0\pm6.5$ & 22\tiny{$^{+6}_{-5}$}      &  4.1 & 5.5 & $--$     & $--$             &6.0    & 0.92   & 0.002     & 0.08     & $--$      \\
 9  &$21.5\pm3.0$    &$51.0\pm5.0$ & 54\tiny{$^{+35}_{-21}$}   & 4.6  & 6.2 & $^{\tiny{5}\,}0.04$      & $0.15$         & 6.8   & 0.90   & 0.004      & 0.08      & 0.02      \\
 11 &$22.5\pm3.2$      &$38.3\pm11$& 66\tiny{$^{+15}_{-12}$} & 6.4   & 8.6 & $^{\tiny{5}\,}0.08$     & $0.2$           & 9.2   & 0.93   & 0.002     & 0.05       & 0.02      \\
 12 &$21.5\pm2.5$     &$33.0\pm8.0$  & 49\tiny{$^{+41}_{-22}$}   & 2.9   & 4.0 & $^{\tiny{5}\,}0.05$     & $0.15$         & 4.8   & 0.81   & 0.005     & 0.15       & 0.03     \\

\hline
\end{tabular}

\parbox{150mm}{
1: Dust mass determined by fitting single-temperature modified blackbody curve.
2: Dust mass determined by fitting the infrared SEDs with \citet{galliano11} model.
3: Cold molecular hydrogen column density calculated for Galactic X factor, X$_{\rm co,20}$ = $2.0\,{\rm cm}^{-2}\,({\rm K\,km\,s}^{-1})^{-1}$.
4: Total gas mass which is the sum of H{\,\sc i}, H$^+$, warm H$_2$ and cold H$_2$ gas masses.
5: From NANTEN CO observations.
}

\end{table*}

\section{Summary}

We report molecular hydrogen emission toward ten ISM regions in the
LMC, observed with the {\it Spitzer} IRS as part of the SAGE-spec
project. For our analysis, the low-resolution infrared spectra of 12 regions were extracted from IRS spectral cubes and integrated
over a circular region with radius 30\,arcsec. All these twelve
regions are either isolated diffuse regions or associated with H{\,\sc
  ii} regions in the LMC (see Fig. \ref{LMCregions}). The pure rotational
 0--0 transitions of H$_2$ were detected in the spectra toward ten
regions. The excitation diagram analysis of the detected H$_2$
transitions gives constraints on the warm H$_2$ gas mass and
temperature. We have only two line detections for most of the regions,
hence we performed single-temperature fit to the excitation
diagram. For three regions a third H$_2$ line is detected, allowing
for a two-temperature fit. In addition to mass and excitation
temperature of the warm H$_2$, we derived mass (within the 30\,arcsec
circular region) of H$^+$ from H$\alpha$ emission; and mass of H{\,\sc
  i} and cold H$_2$ from ancillary H{\,\sc i} and CO data
respectively. These measurements allow us to distinguish the regions
which are diffuse atomic and molecular. We have also determined the
dust temperature and mass from infrared SED fitting, in order to
compare with the warm H$_2$ gas parameters.
\begin{itemize}

\item Our analysis shows that six regions (regions 1, 2, 7, 9, 11, 12)
  are diffuse atomic in nature where H{\,\sc i} is the dominant ISM component
  $>$50\,per\,cent while four other regions (regions 3, 4, 5 and 6) are diffuse molecular, where
  H$_2$ is dominant ($\ge$50\,per\,cent; see Fig. \ref{his}). In both cases, warm H$_2$
  contributes a significant fraction to the total ISM gas, with a mass of
  $5-17$\,per\,cent of the total gas mass and excitation temperature
  $86-137\,{\rm K}$. Interestingly, the
  amount of warm H$_2$ is found to be equally significant in both atomic
  and molecular dominated environments, regardless of the nature of
  the region.

\item For all ten LMC regions where clear H$_2$ excitation is
  detected, a tight correlation between the H$_2$ and 7.9${\,\rm \mu m}$ PAH
  emission is found (R$\sim$0.96). This indicates a PDR excitation,
  caused by photo-electrons ejected from PAHs by UV radiation. The
  surface brightness ratio H$_2$/PAH shows very little deviation ($0.002-0.008$) which agrees
  with the SINGS normal galaxy samples \citep{roussel07}.

\item The surface brightness ratio H$_2$/TIR shows relatively larger
  scatter compared with the SINGS galaxy samples, but a strong
  correlation is observed between H$_2$ and TIR power with a
  correlation coefficient of 0.92. Since the TIR flux is an approximate measure of the
  input energy into the medium, this relation may represent more efficient
  photo-electric heating in the environment where H$_2$ is excited.

\item The ratio H$_2$/24$\,{\rm \mu m}$ is found to decrease with
  an increase in 24$\,{\rm \mu m}$ power. A similar anti-correlation is also
  found for PAH emission. The ratio 7.9${\,\rm \mu m}$/24$\,{\rm
    \mu m}$ decreases with the 24$\,{\rm \mu m}$ power. This could be due to the increased starlight intensity or
  to the preferential destruction of PAHs at higher 24$\,{\rm \mu
    m}$ emission. On the other hand, it may also be possible that the
  24$\,{\rm \mu m}$ and H$_2$ emission arise from physically unrelated
  regions.

\item We examined various tests for correlation of warm H$_2$ mass and
  excitation temperature with the ISM dust parameters (see
  Fig. \ref{corr1}). There is a moderate positive correlation for the H$_2$ mass with the cold dust
  mass. The warm H$_2$ mass fraction tends to positively
  correlate with the cold dust equilibrium temperature or average
  starlight intensity $\Braket{U}$. 

\section{Acknowledgement}

F.K acknowledges funding from Taiwan's Ministry of Science and Technology (MoST) under grants NSC100-2112-M-001-023-MY3 and MOST ISO-2112-M-001-033.
We thank the referee for fruitful comments.
\end{itemize}

\bibliographystyle{mn2e}
\bibliography{mnemonic,LMC_diffuse}

\label{lastpage}
\end{document}